\documentclass[longauth]{aa} 
\usepackage{url}
\usepackage{graphicx}
\usepackage[varg]{txfonts}
\usepackage{hyperref} 
\hypersetup{breaklinks=true,colorlinks=true,linkcolor=blue,citecolor=blue,urlcolor=blue}
\usepackage{natbib,twoopt}
\bibpunct{(}{)}{;}{a}{}{,}
\usepackage{color}
\usepackage{deluxetable}
\usepackage{forloop}

\graphicspath{{figures/}}

\newcommand{\kms}{km\,s$^{-1}$}

\newcommand{\Msun}{M$_\odot$}

\newcommand{\Reff}{R$_\mathrm{e}$}

\def\todo[#1]#2{\noindent {\color{blue} {\bf[#1]:} #2}}

\begin{document} 

\title{Stellar kinematics across the Hubble sequence in the CALIFA survey: General properties and aperture corrections}

\titlerunning{Stellar kinematics across the Hubble sequence}
\authorrunning{Falc{\'o}n-Barroso et al.}

\author{J.~Falc\'on-Barroso\inst{\ref{inst1},\ref{inst2}}\thanks{Email: jfalcon@iac.es}
   \and M.~Lyubenova\inst{\ref{inst3},\ref{inst4}}
   \and G.~van~de~Ven\inst{\ref{inst4}}
   \and J.~Mendez-Abreu\inst{\ref{inst5}}
   \and J.~A.~L.~Aguerri\inst{\ref{inst1},\ref{inst2}}
   \and B.~Garc\'ia-Lorenzo\inst{\ref{inst1},\ref{inst2}}
   \and S.~Bekerait{\.e}\inst{\ref{inst6}}
   \and S.~F.~S\'anchez\inst{\ref{inst7}}
   \and B.~Husemann\inst{\ref{inst8}}
   \and R.~Garc{\'\i}a-Benito\inst{\ref{inst9}}
   \and D.~Mast\inst{\ref{inst10},\ref{inst11}}
   \and C.J.~Walcher\inst{\ref{inst6}}
   \and S.~Zibetti\inst{\ref{inst12}}
   \and J.~K.~Barrera-Ballesteros\inst{\ref{inst13}}
   \and L.~Galbany\inst{\ref{inst14},\ref{inst15}}
   \and P.~S\'anchez-Bl\'azquez\inst{\ref{inst16},\ref{inst17}}
   \and R.~Singh\inst{\ref{inst4}}
   \and R.~C.~E.~van den Bosch\inst{\ref{inst4}}
   \and V.~Wild\inst{\ref{inst5}}
   \and\newline L. Zhu\inst{\ref{inst4}}
   \and J.~Bland-Hawthorn\inst{\ref{inst18}}
   \and R.~Cid~Fernandes\inst{\ref{inst19}}
   \and A.~de~Lorenzo-C{\'a}ceres\inst{\ref{inst5}}
   \and A.~Gallazzi\inst{\ref{inst12}}
   \and R.~M.~Gonz\'alez Delgado \inst{\ref{inst9}}
   \and R.~A.~Marino\inst{\ref{inst20},\ref{inst21}}
   \and I.~M\'arquez\inst{\ref{inst9}}
   \and E.~P\'erez\inst{\ref{inst9}}
   \and I.~P\'erez\inst{\ref{inst22},\ref{inst23}}
   \and M.~M.~Roth\inst{\ref{inst6}}
   \and F.~F.~Rosales-Ortega\inst{\ref{inst24}}
   \and T.~Ruiz-Lara\inst{\ref{inst22}}
   \and L.~Wisotzki\inst{\ref{inst6}}
   \and B.~Ziegler\inst{\ref{inst25}}
   \and the CALIFA collaboration
} 
\institute{
Instituto de Astrof\'isica de Canarias, V\'ia L\'actea s/n, E-38205 La Laguna, Tenerife, Spain\label{inst1}
\and
Departamento de Astrof\'isica, Universidad de La Laguna, E-38205 La Laguna, Tenerife, Spain\label{inst2}
\and 
Kapteyn Astronomical Institute, University of Groningen,Postbus 800, NL-9700 AV Groningen, Netherlands\label{inst3}
\and
Max-Planck-Institut f\"ur Astronomie, K\"onigstuhl 17, D-69117 Heidelberg, Germany\label{inst4}
\and
School of Physics and Astronomy, University of St Andrews, North Haugh, St Andrews, KY16 9SS, UK (SUPA)\label{inst5}
\and
Leibniz-Institut f\"ur Astrophysik Potsdam (AIP), An der Sternwarte 16, D-14482 Potsdam, Germany,\label{inst6}
\and
Instituto de Astronom\'ia, Universidad Nacional Aut\'onoma de M\'exico, Apartado Postal 70-264, M\'exico D.F., 04510 M\'exico\label{inst7}
\and 
European Southern Observatory, Karl-Schwarzschild-Str. 2, 85748 Garching b. M\"unchen, Germany\label{inst8}
\and
Instituto de Astrof\'isica de Andaluc\'ia (IAA/CSIC), Glorieta de la Astronom\'ia s/n Aptdo. 3004, E-18080 Granada, Spain\label{inst9}
\and 
Observatorio Astron\'omico, Laprida 854, X5000BGR, C\'ordoba, Argentina.\label{inst10}
\and 
Consejo de Investigaciones Cient\'{i}ficas y T\'ecnicas de la Rep\'ublica Argentina, Avda. Rivadavia 1917, C1033AAJ, CABA, Argentina.\label{inst11}
\and 
INAF-Osservatorio Astrofisico di Arcetri - Largo Enrico Fermi, 5 - I-50125 Firenze, Italy\label{inst12}
\and
Space Telescope Science Institute, 3700 San Martin Drive, Baltimore, MD 21218, USA\label{inst13}
\and
Millennium Institute of Astrophysics, Chile\label{inst14}
\and
Departamento de Astronom\'ia, Universidad de Chile, Camino El Observatorio 1515, Las Condes, Santiago, Chile\label{inst15}
\and
Departamento de F\'{\i}sica Te\'orica, Universidad Aut\'onoma de Madrid, E-28049, Madrid, Spain\label{inst16}
\and
Instituto de Astrof\'isica, Pontificia Universidad Cat\'olica de Chile, Av. Vicu{\~n}a Mackenna 4860, Santiago , Chile.\label{inst17}
\and
Sydney Institute for Astronomy, School of Physics A28, University of Sydney, NSW 2006, Australia\label{inst18}
\and
Departamento de F\'{\i}sica, Universidade Federal de Santa Catarina, P.O. Box 476, 88040-900, Florian\'opolis, SC, Brazil\label{inst19}
\and
Departamento de Astrof{\'i}sica y CC. de la Atm{\'o}sfera, Universidad Complutense de Madrid, E-28040, Madrid, Spain\label{inst20}
\and
ETH Z\"urich, Institute for Astronomy, Wolfgang-Pauli-Str. 27, 8093 Z\"urich, Switzerland\label{inst21}
\and
Departamento de F\'{\i}sica Te\'orica y del Cosmos, University of Granada, Facultad de Ciencias (Edificio Mecenas), E-18071 Granada, Spain\label{inst22}
\and
Instituto Carlos I de F\'{\i}sica Te\'orica y Computaci\'on\label{inst23}
\and
Instituto Nacional de Astrof{\'i}sica, {\'O}ptica y Electr{\'o}nica, Luis E. Erro 1, 72840 Tonantzintla, Puebla, Mexico\label{inst24}
\and 
University of Vienna, Department of Astrophysics, T\"urkenschanzstr. 17, 1180 Vienna, Austria\label{inst25}
}

\date{Received ...,.. ; accepted ..., ..}

\abstract{We present the stellar kinematic maps of a large sample of galaxies 
from the integral-field spectroscopic survey CALIFA. The sample comprises 300 
galaxies displaying a wide range of morphologies across the Hubble sequence, 
from ellipticals to late-type spirals. This dataset allows us to homogeneously 
extract stellar kinematics up to several effective radii. In this paper, 
we describe the level of completeness of this subset of galaxies with respect to 
the full CALIFA sample, as well as the virtues and limitations of the kinematic 
extraction compared to other well-known integral-field surveys. In addition, we 
provide averaged integrated velocity dispersion radial profiles for different 
galaxy types, which are particularly useful to apply aperture corrections for 
single aperture measurements or poorly resolved stellar kinematics of 
high-redshift sources. The work presented in this paper sets the basis for the 
study of more general properties of galaxies that will be explored in subsequent 
papers of the survey.}


\maketitle


\section{Introduction}
\label{S:intro}
The motion of stars within galaxies is a fundamental property set very
early on in their life. Ever since the detection of rotation of stars in the
Milky Way and nearby systems \citep[e.g.,][]{lindblad1927,mayall1951,munch1960},
the study of stellar motions has been a fruitful avenue to pose important
constraints on our knowledge about galaxy formation and evolution. The analysis
of rotational over random motions in early-type galaxies, for instance, has led to
the realization that bright early-type galaxies are likely triaxial objects
supported by orbital anisotropy \citep[e.g.,][]{bertola75,illingworth77,binney78}, 
rather than rotation.

The coupling of long-slit spectrographs with telescopes 2 to 4\,m in size has 
provided, over the last three decades, a wealth of spatially resolved 
observations that has greatly improved our understanding of the overall stellar 
motion and level of kinematic substructure in external galaxies 
\citep[e.g.,][]{defis83,bertola84,bender94,fisher97,simien97, rubin99, 
vegabeltran01,aguerri03,fb03,pizzella04,remco15}.

While the first integral-field units (IFUs) were already in place in the 
mid-90's \citep[e.g.,][]{bacon95}, the first serious efforts to measure stellar 
kinematics on large samples of galaxies using these kinds of instruments did not 
occur until year 2001. One of the pioneer projects in this respect was the 
SAURON survey \citep{bacon01, dezeeuw02}. With a representative sample of 72 
galaxies (24 ellipticals, 24 lenticulars, and 24 early-type spirals, later 
extended with observations of 18 late-type spirals), this survey has set the 
reference for stellar kinematic IFU studies 
\citep[e.g.,][]{emsellem04,fb06,ganda06}. The discovery of the slow and fast 
rotator families in early-type galaxies \citep{emsellem07} served as the trigger 
for a larger project: the ATLAS$^{\rm 3D}$ survey \citep{cappellari_etal_2011}, 
in which a volume complete sample of 260 early-type galaxies revisited many 
kinematic aspects, from the amount of global angular momentum 
\citep{emsellem_etal_2011} to a detailed account of kinemetric features 
\citep{krajnovic06,krajnovic_etal_2011}. In parallel, the DiskMass survey 
mapped, the stellar kinematic properties of nearby late-type spirals with the 
aid of the PPak IFU \citep{Roth_etal_2005, Kelz_etal_2006}.

The CALIFA survey \citep{Sanchez_etal_2012} was born to fill in existing gaps in 
other IFU surveys and to provide a morphologically unbiased view of the stellar 
kinematics in galaxies based on a large ($\sim$\,600 galaxies) and homogeneous 
integral-field spectroscopic dataset. The main advantage of CALIFA over existing 
surveys resides in a sample selection that includes all  morphological types, as 
well as a field-of-view (FoV) that extends up to several effective radii 
(\Reff). While CALIFA is no longer the IFU survey with the largest number of 
observed objects in the nearby Universe, it still provides the best compromise 
between spatial coverage (1.8--3.7\,\Reff) and sampling ($\sim$1\,kpc). 
Currently ongoing IFU surveys are hampered in one way or another by these 
factors, for example, SAMI covers areas within 1.1--2.9\,\Reff\ with a spatial 
sampling $\sim$\,1.7\,kpc \citep{sami,bryant15}, while MaNGA primary sample 
targets have a spatial sampling of $\sim$3\,kpc within 1.5\,\Reff\ 
\citep{manga}. The real revolution in this respect will take place when MUSE at 
the Very Large Telescope \citep{bacon10} is used in survey mode, as anticipated 
by the very spectacular stellar kinematic cases presented in the first few years 
of operations \citep[e.g.,][]{emsellem14,davor15,gadotti15,iodice15}.\looseness-1

The goal of this paper is to present the first stellar kinematic maps extracted 
from the CALIFA survey, describe all the technical details of the extraction, 
and provide basic stellar velocity dispersion aperture corrections for 
elliptical and spiral galaxies. The maps presented here have already been used 
within the survey to establish the effect of galaxy interactions on the stellar 
kinematics of galaxies \citep{barrera14,barrera15}, constrain the pattern speed 
of barred galaxies across the Hubble sequence \citep{aguerri15}, to present a 
volume-complete Tully-Fisher relation \citep{simona16}, and the velocity 
function of galaxies as a benchmark for numerical simulations \citep{simona16b}. 
Forthcoming papers of the survey will make use of this information, for example, 
to revisit the distribution of global angular momentum in nearby galaxies and 
determine their dark matter content. \citet{fb15} provides a preview of some 
highlights. For results on the kinematics of the ionized gas in CALIFA, see 
\citet{garcialorenzo15}.

The paper is organized as follows. Section~\S\ref{S:sample} describes the sample 
of 300 galaxies used in our study and how this sample compares with the full 
CALIFA sample. Section~\ref{S:setup} summarizes the instrumental setup employed 
during the observations. In section~\ref{S:kinextraction} we provide details of 
our kinematic extraction and comparisons with other major IFU surveys. 
Section~\ref{S:sigma_limit} explains the limit set by our instrumental setup in 
the measurement of stellar velocity dispersions. In 
section~\ref{S:aperture_corr} we provide velocity dispersion aperture 
corrections for elliptical and spiral galaxies. Finally, we summarize our work 
and conclusions in section~\ref{S:conclusions}.

\begin{figure}
\centering
\includegraphics[angle=0,width=\linewidth]{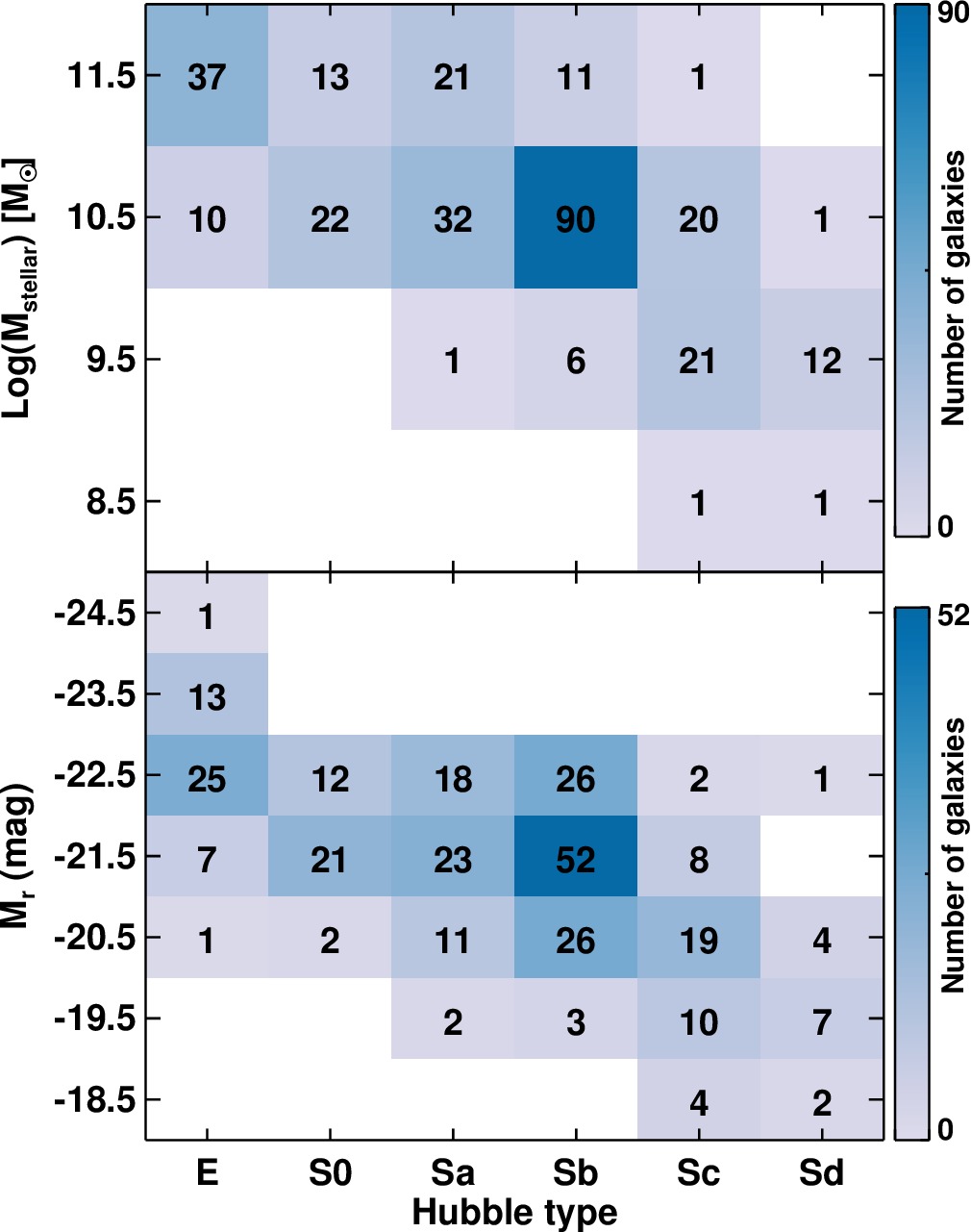}
\caption{Distribution of galaxies in the sample of CALIFA galaxies presented in
this paper (see \S\ref{S:sample}) as a function of Hubble type, stellar
mass, and absolute magnitude in the $r$ band. For convenience, along with the color
bar, we indicate the number of galaxies in each bin.}
\label{fig:sample}
\end{figure}

\section{The CALIFA sample}
\label{S:sample} 

This study is based on observations of 300 galaxies drawn from the 
CALIFA mother and extended samples\footnote{Extended sample galaxies, two objects 
in this study, have CALIFA IDs larger than 1000. See Table~1 and \citet{DR3} as well.}, 
which are part of the photometric catalog of the seventh data release \citep{DR7} 
of the Sloan Digital Sky Survey (SDSS). The main selection criteria in the 
survey is an angular isophotal diameter \hbox{($45\arcsec\leq D_{25} 
\leq80\arcsec$)}, which is followed by a limited range in redshift, 
\hbox{$0.005\leq z \leq0.03$}. These constraints ensure an efficient use of the 
PPak IFU and excludes, together with an apparent magnitude cut at $r-$band 
Petrosian magnitude of $\sim$20\,mag, the presence of too many dwarf galaxies in 
our sample. \citet{Walcher_etal_2014} provides more details 
about the sample selection criteria and an in-depth discussion of the 
effects they introduce in the survey.

The CALIFA sample contains a large number of galaxies with diverse kinematic 
properties: from slow rotating ellipticals, to disk-dominated fast rotating 
galaxies, and perturbed interacting systems. This paper is based on the V1200 
data (see \S\ref{S:setup}) available until June 2014. We removed from our 
original sample of 375 galaxies those cases where the quality of the resulting 
stellar kinematic maps was not sufficient (e.g., poor spatial sampling due to 
low-quality data) to guarantee a meaningful analysis. We also selected out those 
cases whose stellar kinematics appeared highly disturbed by the presence of 
large nearby companions or had clear indications of being in final stages of a 
merging process. While this criteria excluded cases like ARP\,220 (shown in 
Figure~\ref{fig:examples}), it did not remove cases like the Mice galaxies (see 
\citealt{Wild_etal_2014} for a detailed CALIFA study of this system), where the 
interaction has not drastically affected the observed kinematics. 
\citet{barrera14,barrera15} carefully examine the stellar kinematics of merging 
systems in the CALIFA survey. Our final sample thus consists of 300 galaxies.

In Fig.~\ref{fig:sample} we show the distribution of CALIFA galaxies presented 
in this paper as a function of Hubble type, stellar mass, and total absolute 
magnitude in the $r$ band. Hubble type classification was determined after 
a careful visual inspection by several members of the team. Stellar masses and total absolute 
magnitudes were derived following the prescriptions described in 
\citet{Walcher_etal_2014}. Stellar masses assume a Chabrier initial-mass function
\citep{chabrier03}. While the number of galaxies represents a major improvement over 
other integral-field surveys, the selection criteria adopted in the CALIFA 
survey introduce an important shortcoming: the lack of low-mass, low-luminosity 
early-type systems and high-mass, high-luminosity late-type galaxies. 
Another important aspect is that our selection criteria favors edge-on 
orientations for the lowest mass and fainter systems (i.e., Sd galaxies). The 
advantage of this selection, however, is that it allows us to volume-correct 
averaged quantities and thus provide kinematic results that are representative 
of the general population of galaxies. Table~1 contains the basic properties of 
the subset of galaxies of our study.

\begin{figure}
\centering
\includegraphics[angle=0,width=\linewidth]{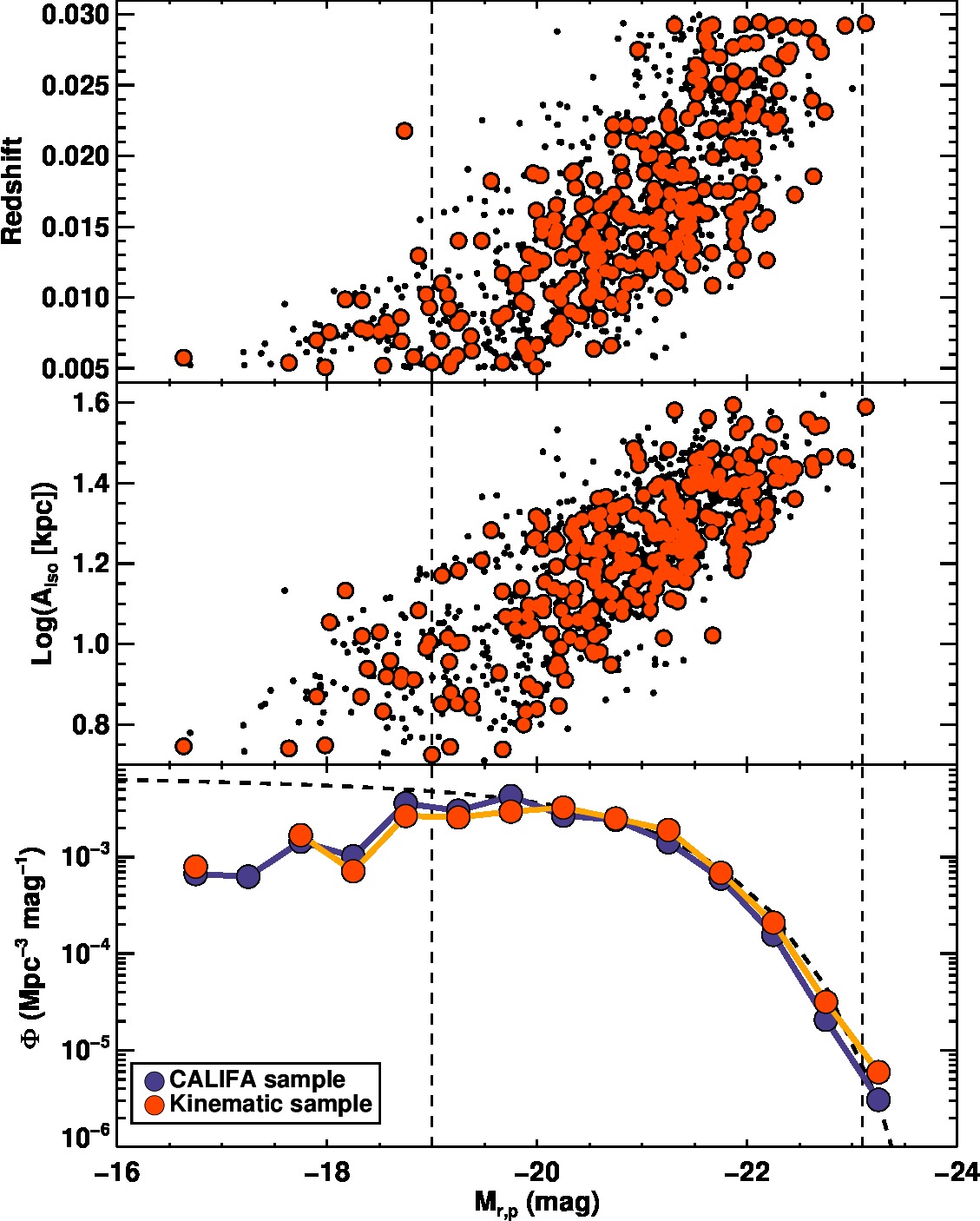}
\caption{(\textit{Top and middle panels}) Distribution of our sample of 300 galaxies 
(orange circles) in redshift, isophotal diameter (A$_{\rm{iso}}$), and absolute 
$r$-band petrosian magnitude (M$_{r\rm{,p}}$). For reference, the CALIFA mother 
sample is shown with black dots. The vertical lines indicate the limits in 
absolute magnitude in which the CALIFA mother sample is representative. (\textit{Bottom panel}) 
Comparison of the luminosity functions of the SDSS (\citealt{blanton03}, thick dashed line), CALIFA 
mother sample (dark blue circles), and the kinematic sample presented here 
(orange circles).}
\label{fig:coverage}
\end{figure}

\begin{figure*}
\centering
\includegraphics[angle=0,width=\textwidth]{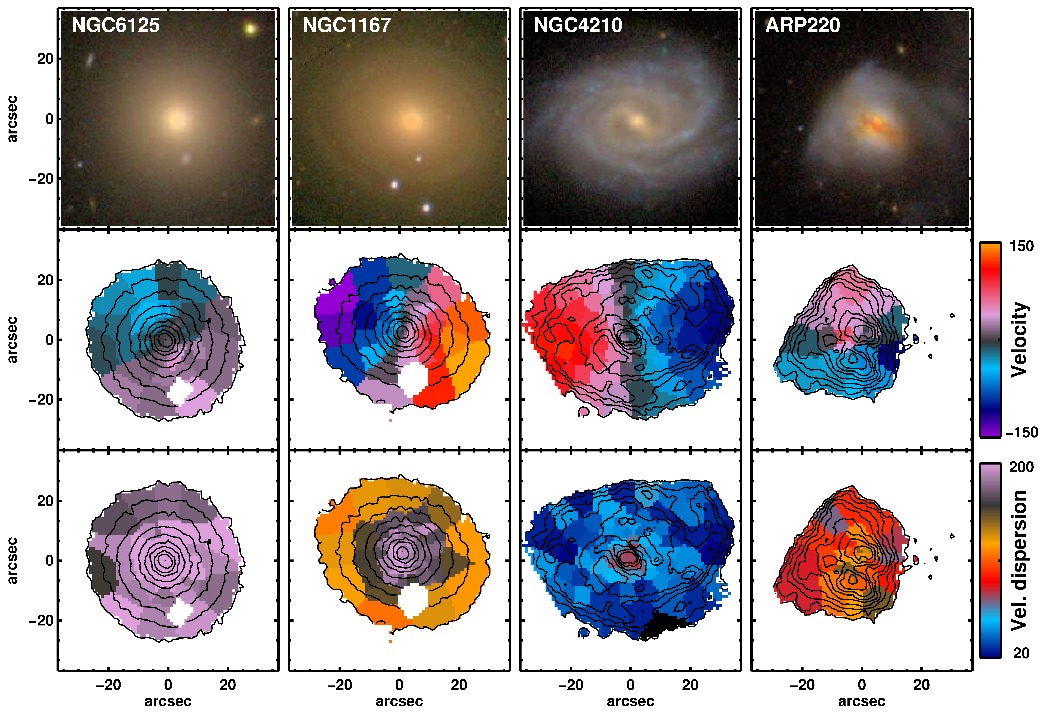}
\caption{Examples of line-of-sight stellar kinematic maps from the CALIFA V1200 
grating dataset. (\textit{Top row}) Color-composite SDSS image of each galaxy. 
(\textit{Middle row}) Stellar velocity maps. (\textit{Bottom row}) Stellar velocity 
dispersion maps. From left to right: \object{NGC\,6125}, a slow-rotator 
elliptical in our sample (i.e., low velocity amplitude and overall large 
velocity dispersion); \object{NGC\,1167}, an early-type spiral galaxy with large 
velocity and central velocity dispersion amplitudes; \object{NGC\,4210}, a 
disk-dominated galaxy (i.e., high velocity amplitude and overall small velocity 
dispersion); \object{ARP\,220}, an interacting system (i.e., with complex 
stellar velocity and velocity dispersions maps). All maps share the same 
velocity and velocity dispersion scale and are in units of \kms\ as indicated in 
the colorbars. Isophotes (black lines) are constructed from the V1200 CALIFA data cube.}
\label{fig:examples}
\end{figure*}

We illustrate how representative our subsample is with respect to the mother 
sample in Fig.~\ref{fig:coverage}. The top and middle panels of the figure show 
the distribution of both the mother sample and our subsample in redshift, 
isophotal diameter (A$_{\rm{iso}}$) and petrosian $r$-band magnitude 
(M$_{r\rm{,p}}$)\footnote{Total absolute magnitudes are used throughout this 
paper, except in Fig.~\ref{fig:coverage} where petrosian magnitudes are employed 
instead for consistency with \citet{Walcher_etal_2014}.}. The vertical lines 
indicate the limits in absolute magnitude in which the CALIFA mother sample is 
representative. In this space of parameters, the distribution of our subsample is 
consistent with that shown by the mother sample. Furthermore our galaxies 
cover all areas sampled by the mother sample. The bottom panel compares the 
luminosity function of SDSS \citep{blanton03}, the CALIFA mother sample, and the 
subset of 300 galaxies of the kinematic sample. We have applied a 
Kolmogorov-Smirnov test to the different parameters and confirm that the 
kinematic and mother samples are statistically consistent. Therefore the set of 
300 galaxies studied in this paper are a good representation of the overall 
population of galaxies of all morphological types in the nearby Universe, within 
the luminosity and size constraints imposed by the CALIFA target selection.

\section{Instrumental setup}
\label{S:setup}

The data presented in this paper is part of the CALIFA survey and as such were 
observed with the PMAS instrument \citep{Roth_etal_2005} in the PPak mode 
\citep{Verheijen_etal_2004, Kelz_etal_2006}, mounted at the 3.5\,m telescope of 
the Calar Alto observatory. For each galaxy, our observations cover the central 
74\arcsec\,$\times$\,64\arcsec\ using a hexagonal fiber bundle. For a detailed 
description of the observations and data reduction, see the 
CALIFA presentation article \citep{Sanchez_etal_2012} and the CALIFA Data 
Release papers 1 and 2 \citep{Husemann_etal_2013,benito15}. The stellar 
kinematics presented in this paper is based on data from the v1.4 data reduction 
pipeline. Here we give a brief overview of the features 
of the observational setup that are relevant to our scientific interests.

The CALIFA survey is conducted in two instrumental setups: a low resolution mode
(V500) with $R$\,$\sim$\,850 at $\sim$\,5000\,\AA\ and a medium resolution mode
(V1200) with $R$\,$\sim$\,1650 at $\sim$\,4500\,\AA. The V500 grating covers a
broad spectral range (3700--7300\,\AA) and includes a number of absorption and
emission features, from the Ca~H+K and [\ion{O}{II}]$\lambda$3727 to H$\beta$
and [\ion{S}{II}]$\lambda$6731 lines. The V1200 grating covers a smaller
spectral window (3400--4750\,\AA). After careful evaluation of the spectral
resolutions of the two gratings we established a value of 6\,\AA\
(FWHM\,$\sim$\,327\,\kms) for the V500 and 2.3\,\AA\ (FWHM\,$\sim$\,169\,\kms)
for the V1200 gratings, respectively \citep[see][]{Husemann_etal_2013}.\looseness-2

\section{Stellar kinematics extraction}
\label{S:kinextraction}

We extracted the stellar kinematics from every galaxy in a uniform way using 
both instrumental setups, i.e., V500 and V1200. Before accomplishing this, we applied 
spatial masks to remove spurious effects such as bad pixels, nearby objects, and/or 
foreground stars. We then logarithmically rebinned the spectra in each data cube 
to conserve a linear step in velocity space. We trimmed the data to contain only 
a useful spectral range: 3800--7000\,\AA\/ for the V500 and 3750--4550\,\AA\/ 
for the V1200 setup. We then selected for future use all spaxels within the 
isophote level where the average signal-to-noise ratio\footnote{We define our 
S/N as the average within the spectral range used in the fitting process.} (S/N) 
was larger than 3. This cut ensured the removal of low-quality spaxels, which 
could introduce undesired systematic effects in our data at low surface 
brightness regimes. The next step was to spatially bin the data cubes to achieve 
an approximately constant S/N of 20 (per pixel). This value allows us to 
conserve a good spatial resolution while still being able to reliably estimate 
the first two moments of the line-of-sight velocity distribution (LOSVD). For 
this step we used the Voronoi 2D binning method of 
\citet{Cappellari_Copin_2003}. Special care was taken in the S/N calculation to 
account for the correlation in the error spectrum of nearby spaxels (see 
\citealt{Husemann_etal_2013} for details).

We measured the stellar kinematics of all galaxies in our sample using the pPXF 
code of \citet{Cappellari_Emsellem_2004}. We used as templates the Indo-U.S. 
spectral library \citep{Valdes_atal_2004} from which we selected $\sim$\,330 
stars that uniformly cover  the parameter space in gravity, metallicity, and 
effective temperature. The careful choice of stellar spectra is crucial to 
minimize template mismatch effects. We confirmed that, using our subset of 
$\sim$300 stars, we could reproduce the same results obtained using the full 
library. A non-negative linear combination of those stellar templates, convolved 
with a Gaussian LOSVD, was fitted to the spectrum of each Voronoi bin. The 
best-fitting parameters were determined by $\chi^2$ minimization in pixel 
space. In the wavelength region covered by CALIFA, there are several emission 
lines that needed to be masked during the fitting process, for example, [\ion{O}{II}], 
[\ion{Ne}{III}], H$\zeta$, H$\epsilon$, [\ion{S}{II}], H$\delta$, 
[\ion{Fe}{II}], H$\gamma$, [\ion{O}{III}], \ion{He}{II}, [\ion{Ar}{IV}], 
H$\beta$, [\ion{N}{I}], \ion{He}{I}, [\ion{O}{I}], [\ion{N}{II}], and H$\alpha$. 
We used a generous band width of 500\,\kms\ around those lines during the 
fitting process. This window was enough to mask the emission in all our 
galaxies. We also masked the regions affected by sky line residuals and the 
sodium doublet at $\sim$5890\,\AA. Additionally, a low-order additive Legendre 
polynomial was included in the fit to account for small differences in the 
continuum shape between the galaxy spectra and the input library. An order 6 
polynomial was the minimum that ensured no large-scale wiggles in the residual 
spectra. In the end, the best-fitting values ($V$ and $\sigma$), and their 
associated uncertainties, were determined as the bi-weight mean and standard 
deviations of a set of 100 Monte Carlo realizations of the fitting. As expected, 
the distribution of best-fitting parameters from the Monte Carlo iterations are 
well-behaved and follow a Gaussian distribution. The bi-weight values measured from 
those distributions agree very well with those obtained from the direct fitting 
of the original spectra. 

In Fig.~\ref{fig:examples} we show a few representative stellar velocity and 
velocity dispersion maps obtained with the V1200 grating. The remaining maps are 
presented in Appendix~\ref{A:kin_maps} of the \textit{Online Material}. The four 
examples shown in the figure illustrate the diversity in the kinematics observed 
in the survey. \object{NGC\,6125}, is a slow-rotator (e.g., low velocity 
amplitude and overall large velocity dispersion). \object{NGC\,1167} is an 
early-type spiral galaxy with large velocity and central velocity dispersion 
amplitudes. \object{NGC\,4210} is a disk-dominated galaxy (e.g., high velocity 
amplitude and overall small velocity dispersion). \object{ARP\,220} is an 
interacting system (e.g., with complex stellar velocity and velocity dispersions 
maps).

\begin{figure}
\centering
\includegraphics[angle=0,width=\linewidth]{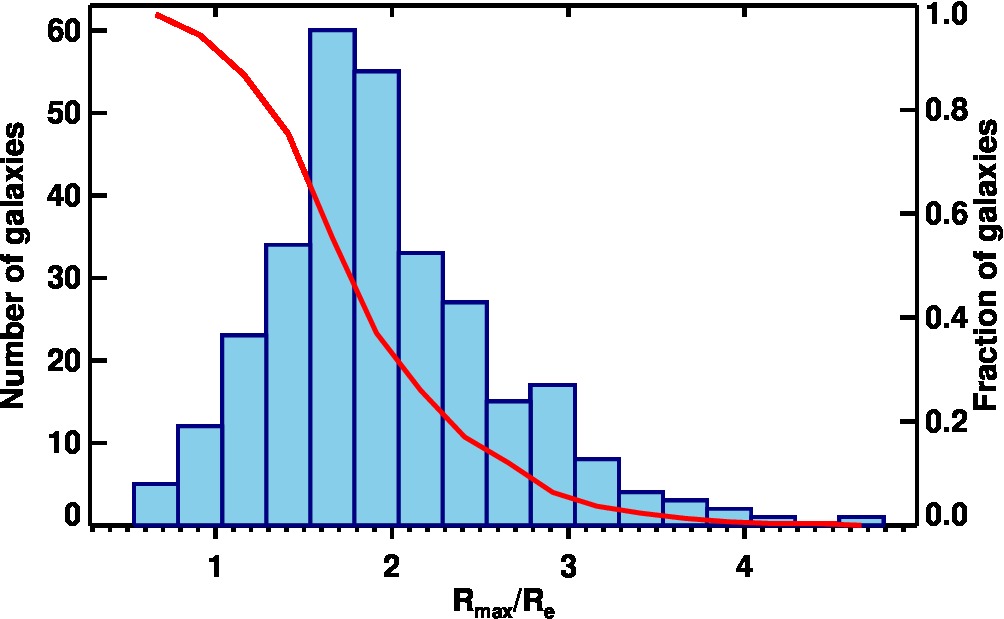}
\caption{Distribution of the radial extent of the CALIFA V1200 stellar kinematics.
The maximum radius reached in our maps is normalized with the semimajor axis
half-light radius (\Reff). The red solid line shows the fraction of galaxies
reaching a certain radial extent, as indicated in the right-hand side vertical axis.}
\label{fig:kin_coverage}
\end{figure}

\subsection{Stellar kinematics coverage}

The CALIFA data presented in this paper allow us to produce stellar kinematic
maps up to a typical surface brightness level of $\sim$\,19\,mag\,arcsec$^{-2}$
(and as faint as 20\,mag\,arcsec$^{-2}$) in $g $ band. We  also quantified
how far, in terms of \Reff, our maps extend. This is shown in
Fig.~\ref{fig:kin_coverage}, where we plot the maximum radius reached by the
measurements in our V1200 maps over \Reff. More than 90\% of the sample covers at
least up to 1\,\Reff\ and 39\% extends beyond 2\,\Reff, with 50\% of the galaxies
reaching at least 1.8\,\Reff. This is a significant improvement over previous
IFU surveys (e.g., SAURON, ATLAS$^{\rm 3D}$), which aimed to probe different properties up
to 1\,\Reff. The strength of those surveys, however, resides in the study of nearby
systems at a higher spatial resolution, which allows them to detect small-scale
inner kinematic subcomponents
\citep[e.g.,][]{mcdermid_etal_2006,krajnovic_etal_2011}.\looseness-1

\subsection{Comparison between V500 and V1200 kinematics}

\begin{figure}
\centering
\includegraphics[angle=0,width=\linewidth]{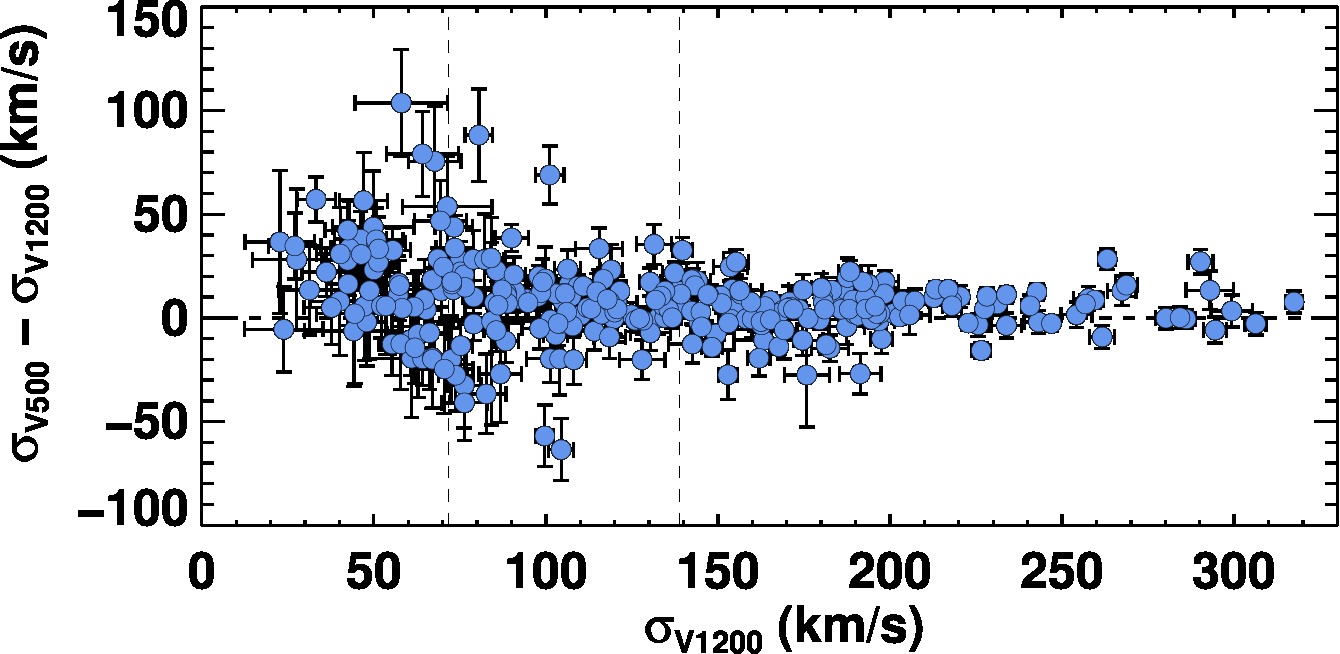}
\caption{Comparison of the stellar velocity dispersion from the CALIFA survey 
for the two instrumental setups: V1200 and V500. The dispersions were 
computed within an aperture of 3\arcsec diameter (i.e., equivalent to the SDSS 
fiber aperture). The vertical dashed lines indicate the spectral resolution of the 
V1200 ($\sigma_{\rm instr.}$\,$\sim$\,72\,\kms) and V500 ($\sigma_{\rm 
instr.}$\,$\sim$\,139\,\kms) setups.}
\label{fig:sigma_comp}
\end{figure}

The two instrumental setups used for the CALIFA survey give us the interesting 
opportunity of measuring the stellar kinematics of galaxies from independent 
datasets. As described in \S\ref{S:setup}, one of these setups (V500) offers a much 
lower spectral resolution than the other, which turns out to be not enough to 
measure the lowest velocity dispersions present in our sample. This issue is 
clearly shown in Fig.~\ref{fig:sigma_comp}, which presents the
difference in velocity dispersion for each setup, measured within a 3\arcsec diameter aperture
centered in each galaxy. In this panel systematic differences
appear at dispersion values below $\sim$100\,\kms. We also compared the 
line-of-sight velocities from each setup (not shown here) and, as expected, 
found that they are well within the uncertainties of our measurements.
Given this limitation, from now on we only report about results coming 
from the V1200 grating.\looseness-2

\subsection{Comparison with other surveys}
\label{SS:compliterature}

\begin{figure}
\centering
\includegraphics[angle=0,width=\linewidth]{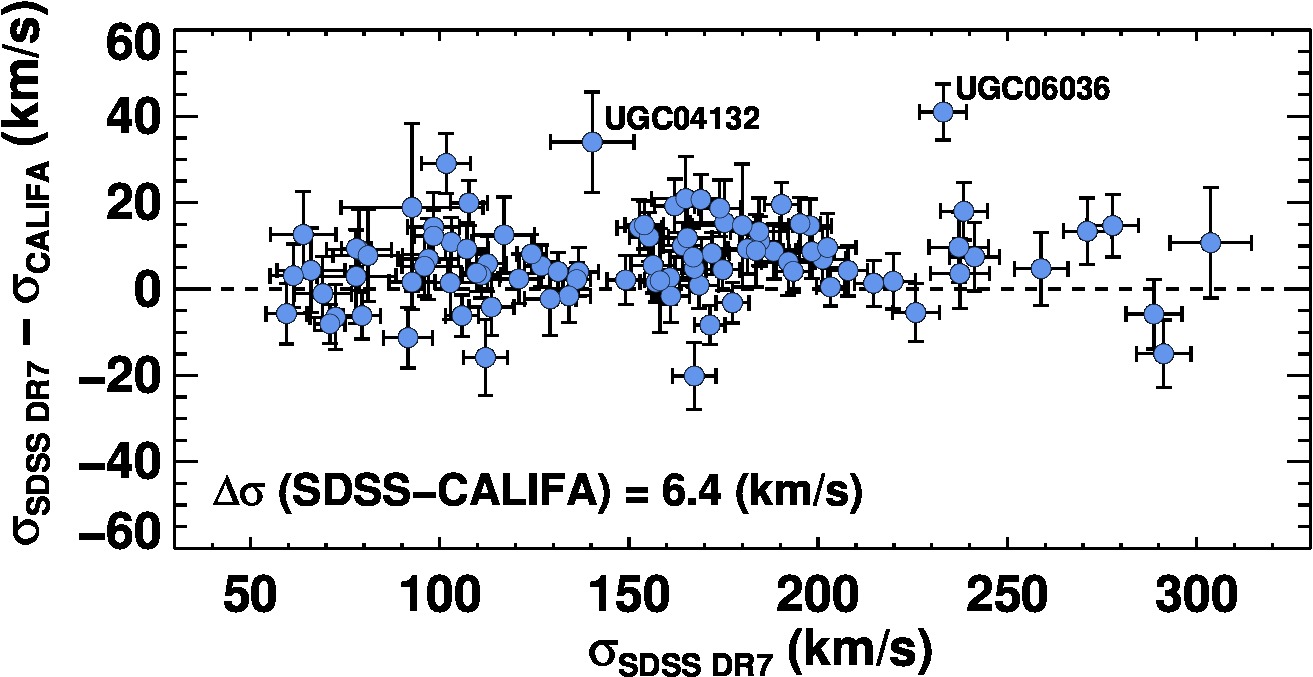}
\caption{Comparison of the stellar velocity dispersion from the CALIFA survey with 
the officially released SDSS DR7 measurements \citep{DR7}. The dispersions were computed within an aperture of 3\arcsec diameter (i.e., equivalent to the 
SDSS fiber aperture).}
\label{fig:sigma_sdss_comp}
\end{figure}

As an additional test to check the reliability and accuracy of our kinematic 
extraction, we compared our central velocity dispersion values with those 
provided by the SDSS DR7 survey \citep{DR7} for those galaxies in our sample 
with SDSS spectroscopy available. We mimicked the SDSS aperture and extracted our 
velocity dispersions within a 3\arcsec\ diameter aperture centered in each galaxy. 
The result of this comparison is presented in Fig.~\ref{fig:sigma_sdss_comp}. 
The agreement between the two sets of measurements is good in general showing 
only a small systematic offset of $\sim6$\,\kms, which is likely due to differences in 
the extraction method, set of templates, point-spread function (PSF)/seeing 
effects, and inaccuracies in the determination of the spectral resolution of 
both the data and templates. Similar levels of discrepancy and systematic 
differences have been identified in the past with SDSS DR7 measurements, even 
using the same SDSS dataset \citep[see Fig.~6 in][]{oh11}.

\begin{figure*}
\centering
\includegraphics[angle=0,width=\linewidth]{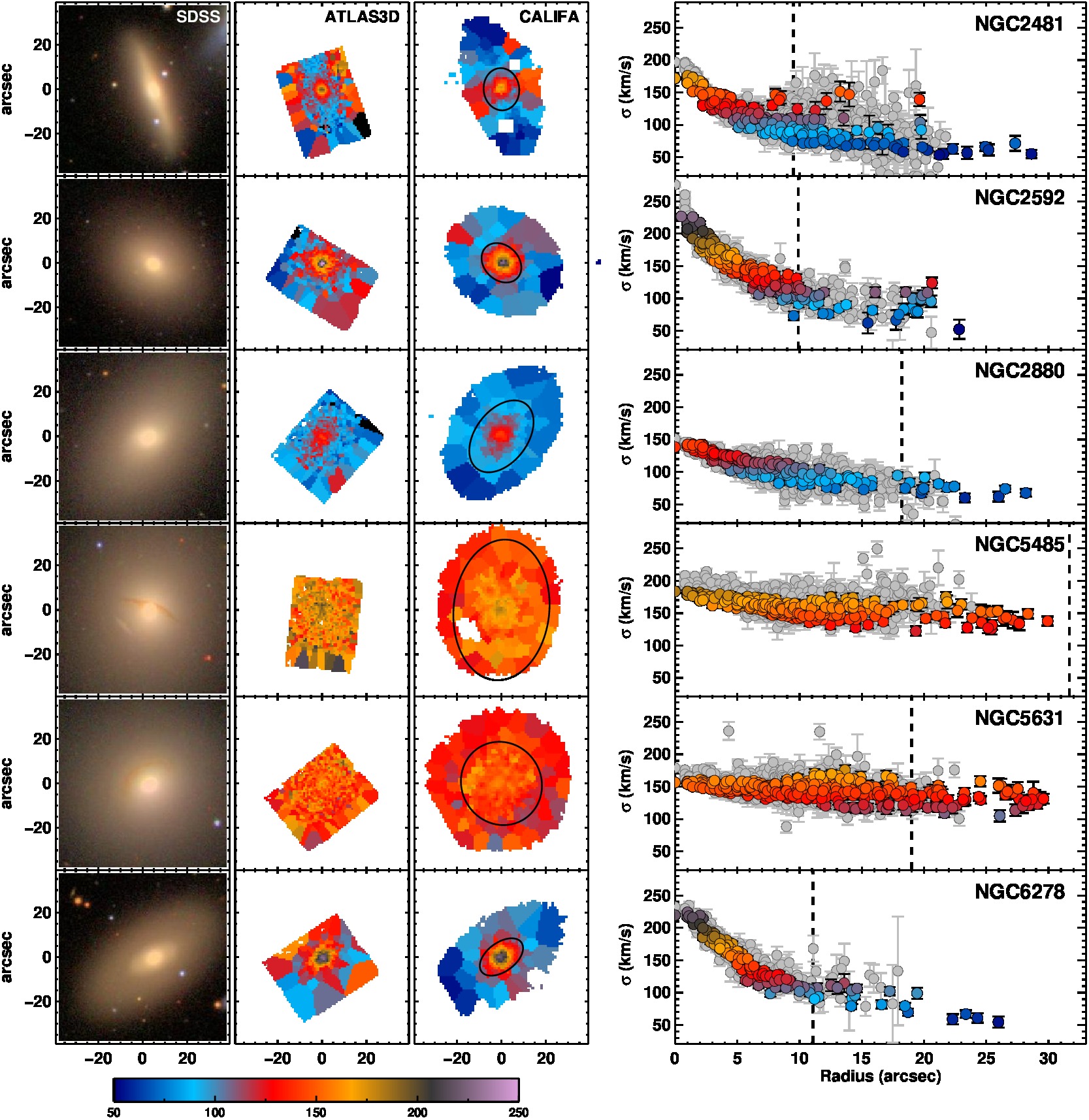}
\caption{Comparison of the stellar velocity dispersion maps and radial profiles 
from the CALIFA and ATLAS$^{\rm 3D}$ surveys. Each row shows (from left to right): a 
color-composite SDSS image of the galaxy, the ATLAS$^{\rm 3D}$ velocity dispersion map, 
the CALIFA velocity dispersion map, and the radial velocity dispersion profile 
(extracted in circular apertures). ATLAS$^{\rm 3D}$ measurements are in gray while the CALIFA 
measurements are in color following the same color scheme of the maps (also indicated 
with the colorbar below). The black ellipse in the maps indicates one 
effective radius in those galaxies. This is also indicated in the radial profile 
panels with a dashed vertical line. Empty (i.e., white) regions within some of 
the CALIFA maps are areas masked during our kinematic extraction. All velocity 
dispersion measurements are expressed in \kms.}
\label{fig:sigma_a3d_comp}
\end{figure*}

\begin{figure*}
\centering
\includegraphics[width=\linewidth]{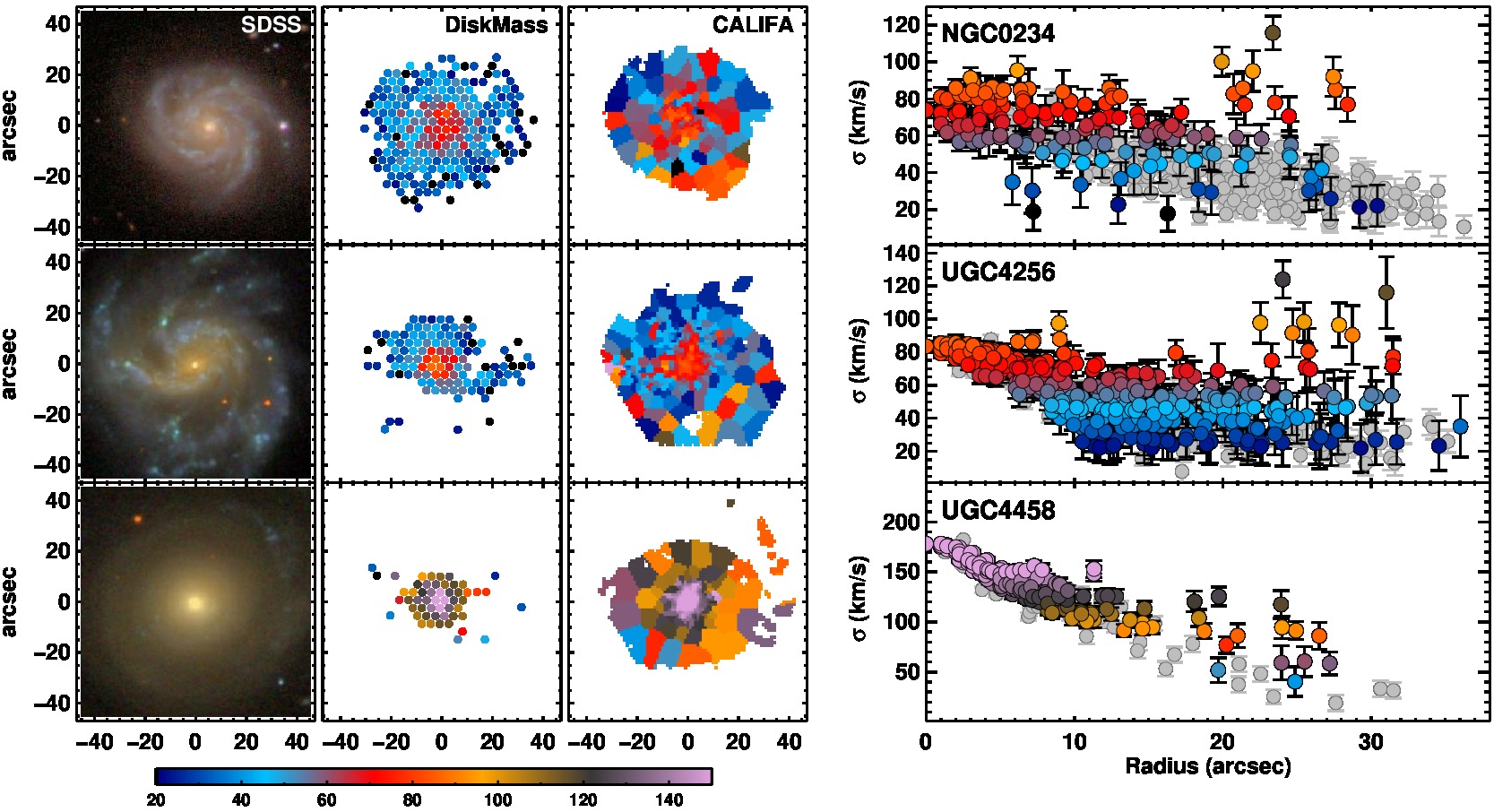}
\caption{Comparison of the stellar velocity dispersion maps and radial profiles 
from the CALIFA and DiskMass surveys. Each row shows (from left to right): a 
color-composite SDSS image of the galaxy, the DiskMass velocity dispersion map 
(see \S\ref{S:sigma_limit} for details), the CALIFA velocity dispersion map, and 
the radial velocity dispersion profile (extracted in circular annuli). 
DiskMass measurements are in gray while the CALIFA measurements are in color following the 
same color scheme of the maps (also indicated with the colorbar below). Empty 
(i.e., white) regions within some of the CALIFA maps are areas masked during our 
kinematic extraction. All velocity dispersion measurements are expressed in 
\kms.}
\label{fig:dms_califa}
\end{figure*}

An even more stringent test is the direct comparison of our stellar velocity 
dispersion maps to those of other surveys. We found up to six objects in 
common with one of the reference IFU surveys today, which is ATLAS$^{\rm 3D}$. We focus 
our test on the velocity dispersion maps, as the velocity maps (not shown here) 
are in good agreement. The results of this comparison are presented in 
Fig.~\ref{fig:sigma_a3d_comp}. By construction, all the ATLAS$^{\rm 3D}$ 
galaxies are early-type systems, which are predominantly red objects with fairly 
high central velocity dispersions. The figure shows a color-composite SDSS image 
of each galaxy in common, as well as the dispersion maps of both surveys and 
radial velocity dispersion profiles (extracted in circular annuli). The overall 
agreement between the two surveys is very good, despite differences in S/N 
thresholds applied in each survey. While the ATLAS$^{\rm 3D}$ data was Voronoi 
binned to a S/N of 40, we deemed it necessary to adopt a threshold S/N of 20 to 
find a good balance between spatial resolution (i.e., Voronoi bin sizes) and 
spatial coverage. The bigger footprint of the PPak IFU allows us to reach well 
beyond 1\,\Reff\ for most of the sample (see Fig.~\ref{fig:coverage}), which is a 
significant improvement over ATLAS$^{\rm 3D}$. The large bins are also 
responsible for the smoother trends observed in the CALIFA radial profiles. 

The only major difference between the two datasets, however, is on the 
central dispersion values. In general, ATLAS$^{\rm 3D}$ values are larger. There 
are two main reasons that could explain this behavior. The PPak IFU is a fiber 
bundle made of 2.7\arcsec\ wide fibers, as opposed to ATLAS$^{\rm 3D}$ with 
$\sim$1\arcsec\ lenslets. While our dithering strategy during observations (see 
\citealt{Sanchez_etal_2012} for details) allowed us to resample our final 
data cube to 1\arcsec\ per spaxel, the original fiber size could result in lower 
velocity dispersion values due to beam smearing. We tested this scenario by 
comparing the central ATLAS$^{\rm 3D}$ values with those obtained by collapsing 
the ATLAS$^{\rm 3D}$ spectra within a 3\arcsec aperture (similar to a CALIFA 
fiber). Our results show that velocity dispersion values can decrease by up to 
15\%. This effect can therefore explain part of the discrepancy between the two surveys. 

In addition, the effective PSF measured for the CALIFA survey 
\citep{benito15} could also affect these values. While reported seeing 
conditions between the two surveys appear similar, if the CALIFA PSF was worse than  the
ATLAS$^{\rm 3D}$ PSF, this could also explain part of the decrease in the central 
velocity dispersion. Based on some simulations carried out in the context of 
another CALIFA paper (\citealt{jairo16}, submitted), we  estimated 
that the PSF can account for up to 5\% difference in the observed values. On top of 
that, the level of Voronoi binning could play a similar role, although this 
seems unlikely in our case as the central spaxels remain mostly 
unbinned.\looseness-2 

\section{Reliability of velocity dispersion measurements below the instrumental resolution}
\label{S:sigma_limit}

An important aspect to consider when extracting stellar kinematics of galaxies 
is to understand the limiting velocity dispersions one can reach given the 
spectral resolution provided by the instrument used. The safest option is to use 
an instrumental setup where the spectral resolution is better than the expected 
values. Under certain circumstances, however, it is possible to push this limit 
and measure velocity dispersions below the nominal threshold imposed by the 
instrument. As shown in \citet{rys_etal_2013}, but see also \citet{gonzalez93} 
and \citet{pedraz02}, a combination of high signal-to-noise and spectral 
sampling of the line spread function \citep[e.g.,][]{koleva_etal_2009} makes it 
possible to overcome, to some extent, this limitation. 

\subsection{Comparison between DiskMass and CALIFA datasets}

The nominal spectral dispersion of the CALIFA V1200 data is $\sigma_{\rm 
intr}$\,$\approx$\,72\,\kms. We determined the velocity dispersion limit of 
our data using as a reference three galaxies in the DiskMass survey 
\citep[DMS;][]{diskmassI}. This dataset was designed to measure velocity 
dispersions in face-on, disk galaxies and the spectral resolution of the 
instrument (FWHM=0.69\AA) was chosen to safely reach values around 
$\sim$\,17\,\kms\ \citep{diskmassVI}. The PPak was custom built for the DMS and  
subsequently employed in the CALIFA 
survey, which can help suppress systematic effects inherent in the analysis. 
The DMS team has kindly provided their data for three galaxies. One was already in 
common with the CALIFA survey (NGC\,234). We observed two more, specifically 
for these tests, in 21--23 February 2014 (UGC\,4256, UGC\,4458), using the same 
V1200 instrumental configuration of the main survey.

Before carrying out our tests, and to account for potential systematic 
effects, we checked that neither the method (cross-correlation technique versus 
pPXF) used to measure the stellar kinematics had a strong impact on the 
resulting velocity dispersions. Our own extraction, using pPXF, of velocity 
dispersions from DMS data provided fully consistent results. The choice of 
templates, whether a single star (as the DMS team used) or a full stellar 
library (like in our case), did not cause any systematic difference in this 
particular exercise. Therefore template mismatch is not an issue in these  
tests. The successful comparison of the two methods using the same data 
was also reported by \citet{diskmassIII}.

In addition to the difference in spectral resolution and template mismatch, there are 
some further differences with respect to the DMS team analysis that can cause 
systematic effects in the velocity dispersion values. Spatial binning is 
desirable in general to reach a threshold S/N, but it can also have the negative 
effect of artificially broadening the line-of-sight velocity distribution. This 
is more acute in the outer regions of galaxies, where the S/N drops quickly and 
the combination of a larger number of spectra is required. Despite this 
drawback, we chose to Voronoi bin the data to ensure a minimum quality of the 
spectra used to derive the velocity dispersion. The DMS team preferred to 
extract their values on single spaxels of $\sim$3\arcsec\ diameter, and remove 
values with uncertainties larger than 8\,\kms (see \S\,7.3.2 of 
\citealt{diskmassVI}).\looseness-1

Another important issue is the wavelength range used to derive the velocity 
dispersion. The DMS values rely on measurements in the short spectral range 
between 4980--5370\,\AA. Our CALIFA values are based on fits between 
3750--4550\,\AA. While a longer baseline is in principle preferred, different spectral features may have slightly distinct 
intrinsic broadening (at the spectral resolutions we are considering here). We 
believe this may be the case in the CALIFA spectral range with the Ca~H+K lines. 
We attempted to derive our stellar kinematics ignoring those lines, but results 
were noisier and uncertainties larger as the fits rely on a few weak spectral 
features, for example, Fe ($\lambda$\,4383\,\AA), H$\gamma$, and H$\delta$. As 
shown in \citet{kb00}, the Ca~H+K lines are reliable features to obtain stellar 
kinematics in all kinds of galaxies, although their results appeared to be more 
uncertain for late-type systems. This may be the culprit of some of the 
differences we see with the DiskMass survey (see below). The detailed 
characterization of all these effects is a complex task, and even if we could 
measure the systematic deviations introduced by each effect, it is not totally 
obvious they would affect different kinds of galaxies in the same manner (e.g., 
emission-free early-type galaxies versus highly star-forming spiral disks). 

\begin{figure}
\centering
\includegraphics[angle=0,width=\linewidth]{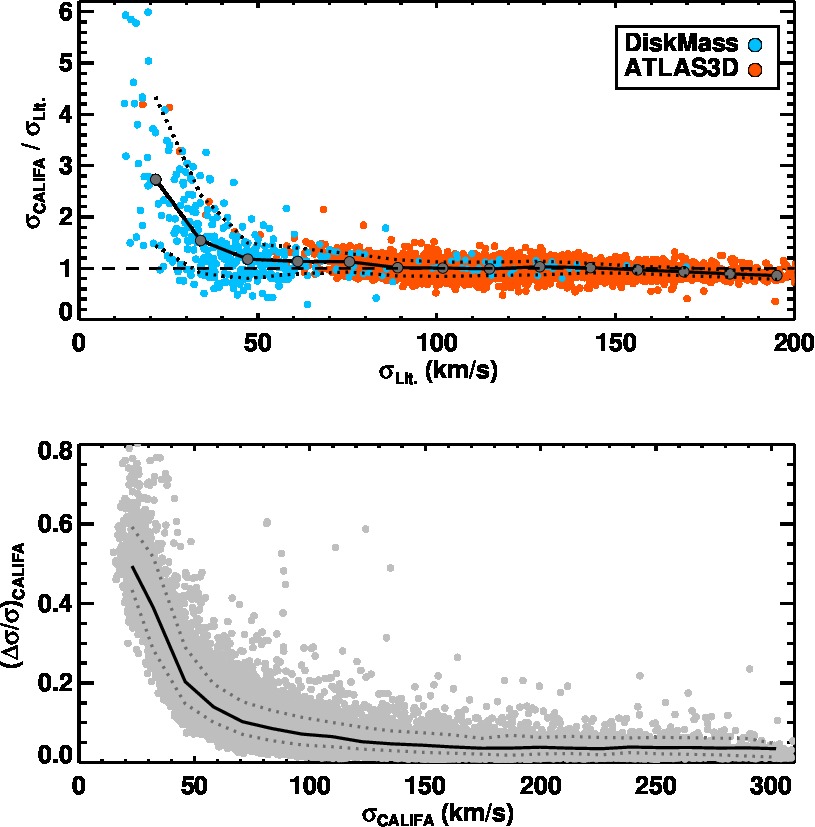}
\caption{Characterization of the biases and relative uncertainties in the 
velocity dispersions of the CALIFA survey. (\textit{Top panel}) Ratio between 
the CALIFA and DMS/ATLAS3D measurements at the locations of the DMS/ATLAS3D 
values (see \S\ref{SS:bias} for details). (\textit{Bottom panel}) Relative 
uncertainties in the velocity dispersion values of the CALIFA survey using the 
Voronoi values and uncertainties for all the galaxies presented here. In both 
panels, the area delimited by the dotted lines indicates the 16\% and 84\% 
percentiles of the distribution of gray points. The solid lines and gray points 
indicate the median of the distributions.}
\label{fig:sigma_bias}
\end{figure}

Figure~\ref{fig:dms_califa} shows the comparison of our own CALIFA data with the 
stellar velocity dispersions measured by the DMS team on the three galaxies in 
common. As in Fig.~\ref{fig:sigma_a3d_comp}, we plot a SDSS color image of each 
galaxy, velocity dispersion maps of the two surveys, and also radial profiles 
(extracted in circular annuli) for a more direct comparison. When displaying the 
maps and radial profiles, we plot the individual spaxel measurements provided by 
the DMS team and our Voronoi binned values. The agreement between the two 
datasets is good overall. We do see discrepancies in some measurements (most 
noticeable in NGC\,234 and UGC\,4256). It appears that some of our CALIFA 
measurements are much larger than those reported by the DMS team at a given 
radius. We explored the reasons for these discrepant values and 
concluded that they occur in low surface brightness regions 
($\mu_{\rm{B}}$\,$\ge$\,22\,mag\,arcsec$^2$) that are often affected by dust or unmasked 
foreground stars. They often have S/N slightly below the nominal S/N=20 
threshold, which is permitted by the Voronoi binning routine within some 
tolerance \cite[see][]{Cappellari_Copin_2003}. These values are also naturally 
associated with large Voronoi bins, which can also artificially help to increase 
the broadening. However, the pPXF fits in those 
regions are not particularly worse than in other areas with similar level of 
binning, S/N, or surface brightness levels. Given that there might be some 
physical insight as to why those values are high (e.g., dust obscuration, 
multiple kinematic components, and kinematic flaring in the outer parts of galaxies), we prefer to keep them in our data release and let the user, based on 
diagnostic parameters we provide, decide whether to include or exclude them 
depending on their science case. This effect is not evident in the DMS values 
owing to the partial field-of-view coverage of their data.\looseness-2 

\subsection{Limiting velocity dispersion and relative uncertainties}
\label{SS:bias}

In order to establish the lowest reliable velocity dispersion we can 
measure, we directly compared  the DMS and ATLAS3D values to our CALIFA 
measurements. This is shown in the top panel of Fig.~\ref{fig:sigma_bias}, where 
we present the ratio of the CALIFA over the DMS and ATLAS3D dispersion values as 
a function of the DMS or ATLAS3D measurements. This exercise determines at 
which velocity dispersion values our CALIFA measurements depart systematically 
from the one-to-one relation. For a fair comparison, we used the Voronoi values 
of our CALIFA maps at the locations of DMS measurements. This is a better 
approach than interpolating our maps at those locations, which may produce 
artifacts. The drawback of this approach is that there is some instrinsic 
scatter produced by the sampling of our points in locations that could be far 
from the Voronoi centroids in our data. It is also sensitive to the different 
levels of scatter of the data points in the surveys (e.g., the scatter of ATLAS3D 
points is larger than CALIFA, see Fig.~\ref{fig:sigma_a3d_comp}). While the 
number of points is not too large for the DMS survey ($\sim$\,360 measurements), 
it is enough to compute some statistics. Besides the individual datapoints, we 
indicate the limiting 16\% and 84\% percentiles of the distribution with dotted 
lines. The median of the distribution is marked with a solid line. The panel 
shows that velocity dispersion values are consistent within the uncertainties 
down to $\sim$\,40\,\kms. Below that point, CALIFA measurements are 
systematically larger up to a factor $\sim$\,3 on average for $\sigma$ values 
around 20\,\kms. On the high velocity dispersion end, values converge 
asymptotically to unity, as expected, except for the most massive systems where we 
suffer the PSF and beam smearing effects discussed in \S\ref{SS:compliterature} 
for ATLAS3D.

In addition to the potential bias in our measurements, it is also interesting to 
determine the relative uncertainties of our measurements at different velocity 
dispersion regimes. This is presented in the bottom panel of 
Fig.~\ref{fig:sigma_bias}. We produced this figure using all the individual 
Voronoi bin measurements and uncertainties for the 300 CALIFA galaxies presented 
here. The shaded region and lines as in the top panel. The figure shows that 
uncertainties are rather small around 5\% for $\sigma$\,$\ge$\,150\,\kms. Below 
that value, relative uncertainties increase up to 50\% for velocity 
dispersions $\sigma$\,$\sim$\,20\,\kms. The median uncertainty at 
$\sigma$\,$\sim$\,40\,\kms, where our measurements start deviating 
systematically from the DMS values, is $\sim$\,20\%.

\begin{figure}
\centering
\includegraphics[angle=0,width=\linewidth]{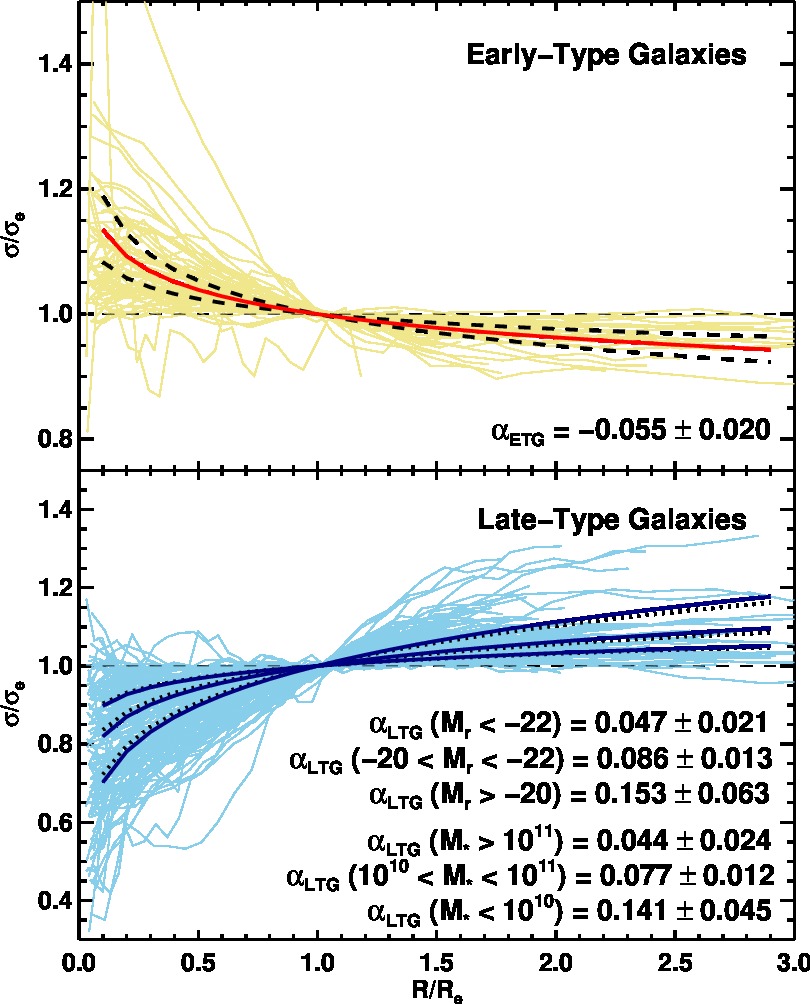}
\caption{Stellar velocity dispersion profiles integrated within elliptical 
apertures with increasing semimajor radius. The profiles are normalized by the 
effective velocity dispersion ($\sigma_{\rm e}$ within the effective radius 
(\Reff). The galaxies were divided depending on the shape of their profile: 
(\textit{top panel}) declining galaxies and (\textit{bottom panel}) steadily 
increasing galaxies, which naturally correspond to early-type and late-type galaxies, 
respectively. For early-type galaxies, the red line is the average fit taking volume corrections 
into account. Dashed lines indicate the uncertainty of the fit. 
For late-type galaxies, dotted lines indicate average fits for different 
intervals of stellar mass, while solid lines indicate average fits for different 
intervals of absolute magnitude. For clarity, we did not include the lines 
with uncertainties in these cases. Averages for late types also take  volume corrections into account.}
\label{fig:sigma_aper}
\end{figure}

\section{Aperture profiles}
\label{S:aperture_corr}

The large number of galaxies across the Hubble sequence in our study allows us 
to estimate velocity dispersion aperture corrections for different groups of 
galaxies. These corrections are useful to homogenize dispersion values measured 
with fiber-fed spectrographs (e.g., SDSS) for galaxies at different distances, 
and they are particularly important for high-redshift studies.

We studied the behavior of the integrated velocity dispersion profiles of our 
galaxies, extracted in elliptical apertures with a fixed position angle and 
ellipticity (as indicated in Table~1). We chose elliptical rather than 
circular apertures to properly account for inclination effects. We used the position angle 
and ellipticity measured in the outer parts of the galaxy (as listed in Table~1).
While this choice ignores potential radial variations in these two parameters 
(e.g., due to bars), the velocity dispersion maps do not appear to be clearly 
influenced by those photometric deviations. This is true in particular for barred 
galaxies, as shown in \citet{seidel15} or \citet{gadotti15}.

We found three types of radial profiles: (\textit{Class~1}) those that decrease 
steadily, (\textit{Class~2}) those decreasing up to a certain radius and then 
increasing again, and (\textit{Class~3}) those that increase steadily with radius. 
We analyzed the type of galaxies conforming each class and found that class~1 
objects are typically early types (e.g., ellipticals, lenticulars and Sa 
galaxies). Class~2 is made of a rather small set of objects ($\sim$20) that are 
mostly early-type galaxies too. They show fairly high dispersions in the center 
but also reasonably high rotation in the outer parts, which drive the increase of 
integrated velocity dispersion at large radii. This effect is even more 
pronounced in class~3 objects, that are predominantly late-type systems (e.g.,  
Sb, Sc, and Sd galaxies).

Given the small number of objects in class~2, we only provide aperture 
corrections for the other two groups (class~1 and~3). We followed previous 
works in the literature and fit the individual profiles of each galaxy in these 
two classes using a power-law function in the form 
\begin{equation}
\left(\frac{\sigma}{\sigma_{\rm e}}\right)=\left(\frac{R}{R_{\rm e}}\right)^\alpha,
\label{eq:aperture}
\end{equation}

\noindent where the effective radius (\Reff) is used as a normalization factor 
for both the radius and velocity dispersion. An important aspect to consider 
during the fitting process was the effect of the PSF in our measurements. We 
account for this effect by convolving our models for each galaxy with the CALIFA PSF 
during the fitting process. As illustrated in the comparison with the 
ATLAS$^{\rm 3D}$ survey data (see \S\,\ref{SS:compliterature}), our velocity 
dispersions are probably smaller than they should at the very center of 
galaxies. Ignoring this effect artificially lowers the $\alpha$ parameter in 
the power-law function. 

Figure~\ref{fig:sigma_aper} shows the individual profiles for classes~1 and~3. 
Class~1, in the top panel, is made of predominantly early-type systems with an 
average stellar mass of $\sim10^{11}$\Msun\ and absolute magnitude M$_{\rm 
r}$\,$\sim$\,$-22$\,mag. We  determined the average profile for the class by 
weighting with the volume correction factor (V$_{\rm max}^{-1}$) of each galaxy. 
That provides a good representation of the average profile for early-type 
galaxies with those properties. The average fit and uncertainty is indicated 
with the solid red line and black dashed lines, respectively. The average value 
of $-0.055$ is in good agreement with corrections reported in previous works 
(e.g., $\alpha=-0.04$, \citealt{jorgensen95}; $\alpha=-0.06$, 
\citealt{mehlert03}; $\alpha=-0.066$, \citealt{cappellari_etal_06}), but see 
\citealt{bb97} for a steeper correction using a different prescription. 

The family of late-type systems in class~3 is much more heterogeneous. We 
decided to divide the sample into three intervals of mass and absolute magnitude. As for early-type galaxies, PSF effects and volume corrections were taken into 
account for the fitting. As illustrated in Fig.~\ref{fig:sigma_aper} (bottom 
panel) there are significant differences in the slopes as a function of mass and 
magnitude. Our results indicate that low-mass and/or low-luminosity spiral 
galaxies display larger $\alpha$ values than high-mass and/or bright systems.

\section{Summary \& conclusions}
\label{S:conclusions} 

In this paper we present stellar kinematic maps for a sample of 300 galaxies 
that are part of the CALIFA survey. The sample covers a wide range of Hubble types, from 
ellipticals to late-spiral galaxies. This subset is a good representation 
of the CALIFA mother sample in terms of redshift, isophotal diameter, and 
absolute magnitude. The large footprint of the PPak IFU, together with the 
average distance of the survey, allow us to measure stellar kinematics well 
beyond 1.8\,\Reff\ for 50\% of the galaxies, reaching out to \mbox{4--5\,\Reff} in a 
few exceptional cases. The penalty, caused by the combination of spatial 
sampling and distance, is the inability to detect kinematically decoupled 
components at the centers of galaxies. Still our data is well suited for the 
study of large-scale kinematic twists or long-axis rotation, which occurs in a 
handful of objects. 

The measurements presented in this paper are in good agreement with those of 
other well-known IFU surveys (e.g., ATLAS$^{\rm 3D}$ and DiskMass). The detailed 
comparison with the DiskMass survey allowed us to establish that we can measure 
reliable velocity dispersion values down to $\sigma$\,$\sim$\,40\,\kms\ (i.e., 
$\sim$\,30\,\kms\ below the instrumental resolution). We also characterized  
the relative uncertainties of our measurements, which are around 5\% for 
$\sigma$\,$\ge$\,150\,\kms. Below that value, relative uncertainties increase 
up to 50\% for velocity dispersions all the way down to $\sigma$\,$\sim$\,20\,\kms.

We also took advantage of our large sample to compute integrated stellar 
velocity dispersion aperture corrections for different sets of galaxies across 
the Hubble sequence. These corrections are particularly useful to homogenize 
dispersion values of galaxies at different distances. We find two main classes 
of integrated aperture radial profiles: steadily decreasing profiles representative of
early-type galaxies, and a second class of systematically increasing profiles typical 
of late-type spiral galaxies. We provide aperture corrections for each class 
for different stellar masses and absolute magnitudes.

The main properties of the sample and the stellar velocity and velocity 
dispersion maps introduced in this paper are available as part of the Online 
Material in Table~1 and Appendix~\ref{A:kin_maps}. The values of the maps 
themselves, together with many diagnostic parameters to assess the quality of 
the measurements, will be made available to the community at the CALIFA website 
(\url{http://califa.caha.es}).

\begin{acknowledgements}
We would like to thank the anonymous referee for constructive comments 
that helped improve some aspects of the original manuscript.
We are also grateful to the DiskMass survey team for sharing their data with us 
for the spectral resolution tests, and to Marc Verheijen and Kyle Westfall 
in particular for in-depth discussions on the topic. 
This study makes use of the data provided by the Calar Alto Legacy Integral 
Field Area ({\sc CALIFA}) survey (http://califa.caha.es). Based on 
observations collected at the Centro Astron\'{o}mico Hispano Alem\'{a}n (CAHA) 
at Calar Alto, operated jointly by the Max-Planck-Institut f\"{u}r Astronomie 
and the Instituto de Astrof\'isica de Andaluc\'ia (CSIC). {\sc CALIFA} is the first 
legacy survey being performed at Calar Alto. The {\sc CALIFA} collaboration 
would like to thank the IAA-CSIC and MPIA-MPG as major partners of the 
observatory, and CAHA itself, for the unique access to telescope time and 
support in manpower and infrastructures.  The {\sc CALIFA} collaboration thanks 
also the CAHA staff for the dedication to this project.\medskip\newline

\textit{Funding and financial support acknowledgements:} J.~F-B. from grant 
AYA2013-48226-C3-1-P from the Spanish Ministry of Economy and Competitiveness 
(MINECO). J.~F-B. and GvdV from the FP7 Marie Curie Actions of the European 
Commission, via the Initial Training Network DAGAL under REA grant agreement 
number 289313.
J.~M-A. and V.~W. acknowledge support from the European Research Council Starting Grant (SEDMorph P.I. V. Wild).
P.S-B acknowledge financial support from the BASAL CATA Center for Astrophysics and Associated Technologies through grant PFB-06.
R.~M.~GD. from grant AYA2014-57490-P.
R.~G-B, R.~M.~GD. and EP acknowledge support from the project JA-FQM-2828.
C.~J.~W. acknowledges support through the Marie Curie Career Integration Grant 303912.
L.~G. from the Ministry of Economy, Development, and Tourism's Millennium Science 
Initiative through grant IC120009 awarded to The Millennium Institute of Astrophysics 
(MAS), and CONICYT through FONDECYT grant 3140566.
I.~M. from grant AYA2013-42227-P.
\end{acknowledgements}

\bibliographystyle{aa}
\bibliography{califakin}


\Online

\begin{deluxetable}{lccccccccc}
\tablewidth{0pt}
\setlength{\tabcolsep}{6pt}
\tablecaption{Basic properties of the CALIFA stellar kinematics sample}
\tablehead{\colhead{Galname} & \colhead{ID} & \colhead{$z$} & \colhead{PA} & \colhead{$\epsilon$} & \colhead{Type} & \colhead{M$_*$} & \colhead{M$_r$} & \colhead{R$_{\rm eff}$} & \colhead{R$_{\rm max}$}
\\
\colhead{~} & \colhead{~} & \colhead{~} & \colhead{(deg)} & \colhead{~} & \colhead{~} & \colhead{(10$^{10}$\,\Msun)} & \colhead{(mag)} & \colhead{(arcsec)} & \colhead{(arcsec)}
}
\startdata
              IC0480 & 159 & 0.015 & 167 & 0.77 & Sc &  1.41 & -20.43 & 24 & 38 \\
              IC0540 & 274 & 0.007 & 170 & 0.63 & Sab &  0.75 & -19.27 & 14 & 27 \\
              IC0674 & 381 & 0.025 & 117 & 0.63 & Sab &  8.09 & -22.07 &  9 & 36 \\
              IC0944 & 663 & 0.023 & 105 & 0.65 & Sab & 18.20 & -22.37 & 19 & 37 \\
              IC1079 & 781 & 0.029 &  82 & 0.51 & E4 & 21.23 & -23.21 & 37 & 26 \\
              IC1151 & 817 & 0.007 &  28 & 0.63 & Scd &  0.70 & -20.26 & 22 & 37 \\
              IC1199 & 824 & 0.016 & 157 & 0.57 & Sb &  4.67 & -21.46 & 20 & 28 \\
              IC1256 & 856 & 0.016 &  89 & 0.36 & Sb &  2.00 & -21.18 & 17 & 25 \\
              IC1528 & 005 & 0.013 &  75 & 0.56 & Sbc &  1.39 & -20.95 & 23 & 41 \\
              IC1652 & 037 & 0.017 & 171 & 0.72 & S0a &  4.09 & -21.20 & 11 & 26 \\
              IC1683 & 043 & 0.016 &  15 & 0.35 & Sb &  3.88 & -21.11 & 13 & 26 \\
              IC1755 & 070 & 0.026 & 155 & 0.75 & Sb &  8.43 & -21.69 & 11 & 31 \\
              IC2101 & 144 & 0.015 & 144 & 0.72 & Scd &  1.74 & -20.81 & 25 & 44 \\
              IC2247 & 186 & 0.014 & 148 & 0.79 & Sab &  3.24 & -20.75 & 21 & 41 \\
              IC2487 & 273 & 0.015 & 162 & 0.79 & Sc &  2.48 & -21.05 & 25 & 40 \\
              IC4566 & 807 & 0.019 & 161 & 0.41 & Sb &  8.99 & -21.96 & 15 & 27 \\
              IC5309 & 906 & 0.014 &  25 & 0.55 & Sc &  1.89 & -20.61 & 17 & 24 \\
              IC5376 & 001 & 0.017 &   3 & 0.69 & Sb &  4.52 & -21.10 & 16 & 36 \\
       MCG-01-54-016 & 878 & 0.010 &  32 & 0.78 & Scd &  0.11 & -18.77 & 24 & 38 \\
       MCG-02-02-030 & 013 & 0.012 & 171 & 0.56 & Sb &  2.34 & -20.88 & 19 & 38 \\
       MCG-02-02-040 & 016 & 0.012 &  53 & 0.47 & Scd &  0.99 & -20.19 & 20 & 34 \\
       MCG-02-03-015 & 032 & 0.019 &  22 & 0.74 & Sab &  4.24 & -21.41 & 12 & 37 \\
       MCG-02-51-004 & 868 & 0.019 & 159 & 0.64 & Sb &  4.79 & -21.69 & 17 & 34 \\
             NGC0001 & 008 & 0.015 & 107 & 0.32 & Sbc &  6.31 & -21.73 & 12 & 30 \\
             NGC0023 & 009 & 0.015 & 177 & 0.30 & Sb & 10.96 & -22.47 & 17 & 26 \\
             NGC0036 & 010 & 0.020 &  24 & 0.48 & Sb &  7.87 & -22.33 & 21 & 33 \\
             NGC0155 & 018 & 0.021 & 167 & 0.14 & E1 & 15.00 & -22.41 & 15 & 25 \\
             NGC0160 & 020 & 0.018 &  49 & 0.47 & Sa & 10.72 & -22.18 & 22 & 35 \\
             NGC0169 & 022 & 0.015 &  90 & 0.47 & Sab & 39.90 & -21.87 & 34 & 34 \\
             NGC0171 & 023 & 0.013 &  32 & 0.05 & Sb &  5.26 & -21.84 & 26 & 32 \\
             NGC0177 & 024 & 0.013 &   8 & 0.42 & Sab &  2.34 & -20.70 & 13 & 37 \\
             NGC0180 & 025 & 0.018 & 167 & 0.34 & Sb &  8.36 & -22.31 & 28 & 41 \\
             NGC0192 & 026 & 0.014 & 168 & 0.57 & Sab &  7.05 & -21.59 & 22 & 39 \\
             NGC0214 & 028 & 0.015 &  50 & 0.26 & Sbc &  6.65 & -22.16 & 18 & 31 \\
             NGC0216 & 027 & 0.005 &  25 & 0.71 & Sd &  0.19 & -18.99 & 20 & 35 \\
             NGC0217 & 029 & 0.013 & 112 & 0.74 & Sa & 12.50 & -21.90 & 23 & 41 \\
             NGC0234 & 031 & 0.015 &  64 & 0.20 & Sc &  4.50 & -21.91 & 20 & 34 \\
             NGC0237 & 030 & 0.014 & 175 & 0.32 & Sc &  2.04 & -21.14 & 15 & 35 \\
             NGC0257 & 033 & 0.018 &  88 & 0.36 & Sc &  6.22 & -22.15 & 21 & 40 \\
             NGC0364 & 035 & 0.017 &  35 & 0.28 & E7 &  9.16 & -21.56 & 15 & 24 \\
             NGC0429 & 036 & 0.019 &  15 & 0.78 & Sa &  6.22 & -21.26 &  6 & 29 \\
             NGC0444 & 039 & 0.016 & 158 & 0.74 & Scd &  0.74 & -20.23 & 23 & 32 \\
             NGC0447 & 038 & 0.019 &  74 & 0.13 & Sa & 13.52 & -22.40 & 28 & 31 \\
             NGC0477 & 042 & 0.020 & 128 & 0.50 & Sbc &  3.14 & -21.69 & 21 & 45 \\
             NGC0496 & 045 & 0.020 &  32 & 0.46 & Scd &  2.59 & -21.40 & 19 & 35 \\
             NGC0499 & 044 & 0.015 &  72 & 0.33 & E5 & 25.18 & -22.48 & 21 & 31 \\
             NGC0504 & 046 & 0.014 &  44 & 0.60 & S0 &  2.95 & -20.76 &  8 & 30 \\
             NGC0517 & 047 & 0.014 &  24 & 0.49 & S0 &  6.64 & -21.35 & 10 & 34 \\
             NGC0528 & 050 & 0.016 &  57 & 0.52 & S0 &  7.48 & -21.68 & 12 & 28 \\
             NGC0529 & 051 & 0.016 &  13 & 0.09 & E4 & 12.25 & -22.27 & 12 & 37 \\
             NGC0551 & 052 & 0.017 & 137 & 0.56 & Sbc &  4.38 & -21.52 & 19 & 44 \\
             NGC0681 & 061 & 0.006 &  65 & 0.33 & Sa &  3.10 & -20.71 & 30 & 37 \\
             NGC0741 & 068 & 0.019 &  85 & 0.22 & E1 & 32.89 & -23.47 & 35 & 32 \\
             NGC0755 & 069 & 0.005 &  49 & 0.61 & Scd &  0.24 & -19.43 & 28 & 39 \\
             NGC0768 & 071 & 0.023 &  33 & 0.61 & Sc &  3.48 & -21.78 & 15 & 34 \\
             NGC0774 & 072 & 0.015 & 164 & 0.18 & S0 &  8.39 & -21.55 & 12 & 26 \\
             NGC0776 & 073 & 0.016 &  41 & 0.10 & Sb &  4.94 & -21.82 & 19 & 32 \\
             NGC0781 & 074 & 0.012 &  11 & 0.70 & Sa &  2.96 & -20.80 &  8 & 32 \\
             NGC0810 & 076 & 0.026 &  27 & 0.34 & E5 & 35.65 & -22.84 & 17 & 20 \\
             NGC0825 & 077 & 0.011 &  50 & 0.51 & Sa &  2.64 & -20.70 & 12 & 34 \\
             NGC0932 & 087 & 0.014 &  65 & 0.08 & S0a &  9.20 & -22.10 & 18 & 33 \\
             NGC1056 & 100 & 0.005 & 153 & 0.32 & Sa &  1.05 & -19.94 & 14 & 37 \\
             NGC1060 & 101 & 0.017 &  75 & 0.18 & E3 & 70.15 & -23.62 & 27 & 26 \\
             NGC1093 & 108 & 0.018 &  99 & 0.39 & Sbc &  3.25 & -21.49 & 13 & 36 \\
             NGC1167 & 119 & 0.016 &  62 & 0.23 & S0 & 49.20 & -22.98 & 24 & 30 \\
             NGC1349 & 127 & 0.022 &  50 & 0.12 & E6 &  8.47 & -22.44 & 17 & 21 \\
             NGC1542 & 131 & 0.012 & 131 & 0.59 & Sab &  2.74 & -20.74 & 15 & 23 \\
             NGC1645 & 134 & 0.016 &  84 & 0.57 & S0a &  6.78 & -21.81 & 13 & 39 \\
             NGC1677 & 143 & 0.009 & 137 & 0.71 & Scd &  0.38 & -19.46 & 12 & 29 \\
             NGC2253 & 147 & 0.012 & 109 & 0.32 & Sbc &  3.34 & -21.55 & 15 & 36 \\
             NGC2347 & 149 & 0.015 &   9 & 0.36 & Sbc &  8.71 & -22.12 & 18 & 42 \\
             NGC2410 & 151 & 0.016 &  34 & 0.68 & Sb &  7.62 & -21.86 & 21 & 37 \\
             NGC2449 & 156 & 0.016 & 135 & 0.52 & Sab &  7.28 & -21.68 & 16 & 33 \\
             NGC2476 & 160 & 0.012 & 136 & 0.29 & E6 &  6.32 & -21.58 &  9 & 22 \\
             NGC2480 & 161 & 0.008 & 167 & 0.43 & Sdm &  0.83 & -19.75 & 35 & 25 \\
             NGC2481 & 162 & 0.007 &   6 & 0.16 & S0 &  4.83 & -20.80 &  9 & 34 \\
             NGC2486 & 163 & 0.015 &  92 & 0.44 & Sab &  3.96 & -21.30 & 15 & 29 \\
             NGC2487 & 164 & 0.016 & 132 & 0.15 & Sb &  5.90 & -22.19 & 28 & 35 \\
             NGC2513 & 171 & 0.016 & 174 & 0.27 & E2 & 34.59 & -22.86 & 26 & 32 \\
             NGC2540 & 183 & 0.021 & 131 & 0.39 & Sbc &  3.32 & -21.62 & 14 & 33 \\
             NGC2553 & 188 & 0.016 &  67 & 0.50 & Sb &  6.89 & -21.30 &  9 & 19 \\
             NGC2554 & 189 & 0.014 & 160 & 0.19 & S0a & 16.33 & -22.59 & 19 & 38 \\
             NGC2592 & 201 & 0.007 &  45 & 0.22 & E4 &  4.15 & -20.72 &  9 & 28 \\
             NGC2604 & 209 & 0.007 &  48 & 0.12 & Sd &  0.46 & -20.24 & 26 & 36 \\
             NGC2639 & 219 & 0.011 & 130 & 0.35 & Sa & 14.72 & -22.33 & 17 & 38 \\
             NGC2730 & 232 & 0.013 &  80 & 0.12 & Scd &  1.31 & -20.94 & 24 & 39 \\
             NGC2880 & 272 & 0.005 & 142 & 0.36 & E7 &  4.69 & -21.10 & 18 & 36 \\
             NGC2906 & 275 & 0.007 &  82 & 0.44 & Sbc &  2.46 & -20.79 & 19 & 33 \\
             NGC2916 & 277 & 0.012 &  19 & 0.36 & Sbc &  5.66 & -22.09 & 26 & 40 \\
             NGC2918 & 279 & 0.023 &  75 & 0.31 & E6 & 27.73 & -22.78 & 12 & 28 \\
             NGC3057 & 312 & 0.005 &  23 & 0.27 & Sdm &  0.12 & -19.17 & 32 & 34 \\
             NGC3106 & 311 & 0.021 & 116 & 0.10 & Sab & 16.29 & -22.79 & 21 & 32 \\
             NGC3158 & 318 & 0.023 & 165 & 0.19 & E3 & 54.70 & -23.70 & 32 & 32 \\
             NGC3160 & 319 & 0.023 & 140 & 0.76 & Sab &  8.99 & -21.51 & 15 & 36 \\
             NGC3300 & 339 & 0.010 & 173 & 0.46 & S0a &  5.78 & -21.41 & 13 & 32 \\
             NGC3303 & 340 & 0.020 & 159 & 0.51 & S0a & 11.51 & -22.33 & 15 & 21 \\
             NGC3381 & 353 & 0.005 &  43 & 0.14 & Sd &  0.48 & -20.08 & 24 & 42 \\
             NGC3615 & 387 & 0.022 &  42 & 0.42 & E5 & 24.15 & -22.98 & 15 & 18 \\
             NGC3687 & 414 & 0.008 & 151 & 0.06 & Sb &  1.88 & -20.97 & 17 & 30 \\
             NGC3811 & 436 & 0.010 & 171 & 0.23 & Sbc &  2.65 & -21.40 & 21 & 39 \\
             NGC3815 & 437 & 0.012 &  67 & 0.50 & Sbc &  2.25 & -21.05 & 14 & 34 \\
             NGC3994 & 476 & 0.010 &   8 & 0.49 & Sbc &  2.65 & -21.22 &  9 & 26 \\
             NGC4003 & 479 & 0.022 & 168 & 0.28 & S0a & 11.83 & -22.00 & 14 & 22 \\
             NGC4047 & 489 & 0.011 &  97 & 0.26 & Sbc &  4.86 & -21.90 & 16 & 33 \\
             NGC4149 & 502 & 0.010 &  85 & 0.60 & Sa &  2.30 & -20.63 & 18 & 36 \\
             NGC4185 & 515 & 0.013 & 164 & 0.33 & Sbc &  4.69 & -21.88 & 30 & 38 \\
             NGC4210 & 518 & 0.009 &  97 & 0.24 & Sb &  1.93 & -20.98 & 21 & 36 \\
             NGC4470 & 548 & 0.008 & 179 & 0.32 & Sc &  0.98 & -20.72 & 15 & 33 \\
             NGC4644 & 569 & 0.016 &  57 & 0.71 & Sb &  2.82 & -21.03 & 12 & 29 \\
            NGC4676A & 577 & 0.022 &   2 & 0.85 & Sdm &  6.50 & -22.17 & 38 & 31 \\
            NGC4676B & 2999 & 0.022 &  43 & 0.44 & S0 &  7.18 & -22.09 & 15 & 23 \\
             NGC4711 & 580 & 0.014 &  41 & 0.47 & Sbc &  2.05 & -21.05 & 17 & 33 \\
             NGC4816 & 588 & 0.023 &  80 & 0.31 & E1 & 32.06 & -23.03 & 30 & 30 \\
            NGC4841A & 589 & 0.023 &  42 & 0.11 & E3 & 35.16 & -22.83 & 20 & 25 \\
             NGC4874 & 592 & 0.024 &  46 & 0.23 & E0 & 49.54 & -24.11 & 55 & 29 \\
             NGC4956 & 602 & 0.016 &  39 & 0.17 & E1 &  9.68 & -22.38 &  9 & 21 \\
             NGC4961 & 603 & 0.009 & 100 & 0.31 & Scd &  0.48 & -20.25 & 15 & 33 \\
             NGC5000 & 608 & 0.019 &   1 & 0.24 & Sbc &  5.37 & -21.81 & 16 & 30 \\
             NGC5016 & 611 & 0.009 &  57 & 0.23 & Sbc &  1.72 & -21.06 & 17 & 35 \\
             NGC5029 & 612 & 0.029 & 149 & 0.40 & E6 & 31.77 & -23.28 & 25 & 28 \\
             NGC5056 & 614 & 0.019 &   3 & 0.44 & Sc &  3.02 & -21.82 & 15 & 38 \\
             NGC5205 & 630 & 0.006 & 169 & 0.35 & Sbc &  0.73 & -20.12 & 19 & 41 \\
             NGC5216 & 633 & 0.010 &  33 & 0.32 & E0 &  3.20 & -21.07 & 20 & 26 \\
             NGC5218 & 634 & 0.010 & 101 & 0.14 & Sab &  4.49 & -21.43 & 18 & 36 \\
             NGC5378 & 676 & 0.010 &  86 & 0.22 & Sb &  3.83 & -21.27 & 24 & 34 \\
             NGC5406 & 684 & 0.018 & 111 & 0.29 & Sb & 18.75 & -22.57 & 20 & 40 \\
             NGC5480 & 707 & 0.006 &  41 & 0.18 & Scd &  1.38 & -20.76 & 25 & 41 \\
             NGC5485 & 708 & 0.006 & 174 & 0.32 & E5 & 10.57 & -21.95 & 31 & 38 \\
             NGC5520 & 715 & 0.006 &  63 & 0.49 & Sbc &  0.73 & -20.18 & 12 & 34 \\
             NGC5614 & 740 & 0.013 & 128 & 0.19 & Sa & 19.86 & -22.77 & 18 & 35 \\
             NGC5630 & 749 & 0.009 &  93 & 0.70 & Sdm &  0.47 & -20.37 & 22 & 38 \\
             NGC5631 & 744 & 0.007 &  30 & 0.06 & S0 &  8.47 & -21.74 & 19 & 34 \\
             NGC5633 & 748 & 0.008 &  16 & 0.26 & Sbc &  1.82 & -20.94 & 13 & 35 \\
             NGC5657 & 754 & 0.013 & 164 & 0.63 & Sbc &  1.92 & -20.98 & 10 & 39 \\
             NGC5682 & 758 & 0.008 & 125 & 0.76 & Scd &  0.25 & -19.39 & 26 & 38 \\
             NGC5720 & 764 & 0.026 & 131 & 0.44 & Sbc &  7.05 & -22.29 & 16 & 27 \\
             NGC5732 & 768 & 0.013 &  43 & 0.48 & Sbc &  0.85 & -20.46 & 14 & 32 \\
             NGC5784 & 778 & 0.018 &  19 & 0.13 & S0 & 16.44 & -22.61 & 13 & 29 \\
             NGC5797 & 780 & 0.013 & 130 & 0.45 & E7 &  7.01 & -22.13 & 18 & 31 \\
             NGC5876 & 787 & 0.011 &  51 & 0.59 & S0a &  7.96 & -21.41 & 12 & 31 \\
             NGC5888 & 789 & 0.029 & 150 & 0.38 & Sb & 16.07 & -22.74 & 16 & 31 \\
             NGC5908 & 791 & 0.011 & 154 & 0.36 & Sa & 16.71 & -22.17 & 34 & 42 \\
             NGC5930 & 795 & 0.009 & 161 & 0.54 & Sab &  4.30 & -21.36 & 16 & 37 \\
             NGC5934 & 796 & 0.019 &  24 & 0.59 & Sb &  8.75 & -21.80 & 13 & 36 \\
             NGC5947 & 4034 & 0.020 &  61 & 0.15 & Sbc &  3.48 & -21.56 & 13 & 32 \\
             NGC5953 & 801 & 0.007 &  43 & 0.10 & Sa &  3.01 & -21.09 & 10 & 34 \\
             NGC5966 & 806 & 0.015 &  83 & 0.39 & E4 & 10.21 & -22.08 & 18 & 34 \\
             NGC5971 & 804 & 0.011 & 132 & 0.56 & Sb &  2.07 & -20.80 & 12 & 27 \\
             NGC5980 & 810 & 0.014 &  11 & 0.60 & Sbc &  5.25 & -21.81 & 17 & 40 \\
             NGC5987 & 809 & 0.010 &  62 & 0.65 & Sa & 16.22 & -22.15 & 33 & 37 \\
             NGC6004 & 813 & 0.013 &  91 & 0.20 & Sbc &  4.86 & -21.86 & 22 & 37 \\
             NGC6020 & 815 & 0.014 & 133 & 0.31 & E4 & 10.02 & -22.08 & 19 & 25 \\
             NGC6021 & 816 & 0.016 & 157 & 0.27 & E5 & 10.14 & -21.88 &  9 & 27 \\
             NGC6032 & 820 & 0.014 &   0 & 0.38 & Sbc &  3.37 & -21.30 & 27 & 39 \\
             NGC6060 & 821 & 0.015 & 102 & 0.57 & Sb &  8.59 & -22.23 & 28 & 35 \\
             NGC6063 & 823 & 0.010 & 156 & 0.44 & Sbc &  1.38 & -20.55 & 20 & 36 \\
             NGC6081 & 826 & 0.017 & 128 & 0.59 & S0a & 13.12 & -21.95 & 12 & 30 \\
             NGC6125 & 829 & 0.015 &   4 & 0.04 & E1 & 24.21 & -22.86 & 21 & 28 \\
             NGC6132 & 831 & 0.017 & 125 & 0.64 & Sbc &  1.63 & -21.04 & 14 & 31 \\
             NGC6146 & 832 & 0.029 &  73 & 0.24 & E5 & 42.56 & -23.48 & 15 & 26 \\
             NGC6150 & 835 & 0.029 &  58 & 0.45 & E7 & 26.67 & -22.65 & 11 & 29 \\
             NGC6168 & 841 & 0.009 & 110 & 0.77 & Sc &  0.73 & -20.00 & 26 & 34 \\
             NGC6173 & 840 & 0.029 & 144 & 0.37 & E6 & 53.09 & -23.85 & 38 & 32 \\
             NGC6186 & 842 & 0.010 &  49 & 0.23 & Sb &  3.71 & -21.24 & 20 & 35 \\
             NGC6278 & 844 & 0.009 & 126 & 0.42 & S0a &  8.30 & -21.49 & 11 & 33 \\
             NGC6301 & 849 & 0.028 & 108 & 0.40 & Sbc & 10.42 & -22.76 & 24 & 39 \\
             NGC6310 & 848 & 0.011 &  69 & 0.72 & Sb &  3.64 & -20.99 & 23 & 33 \\
             NGC6314 & 850 & 0.022 & 173 & 0.47 & Sab & 16.26 & -22.46 & 12 & 37 \\
             NGC6338 & 851 & 0.027 &  15 & 0.38 & E5 & 49.09 & -23.48 & 28 & 26 \\
             NGC6394 & 857 & 0.028 &  42 & 0.64 & Sbc &  7.87 & -21.87 & 14 & 24 \\
             NGC6411 & 859 & 0.012 &  65 & 0.35 & E4 & 12.08 & -22.42 & 34 & 33 \\
             NGC6427 & 860 & 0.011 &  34 & 0.57 & S0 &  5.64 & -21.37 &  8 & 34 \\
             NGC6478 & 862 & 0.023 &  34 & 0.63 & Sc & 10.33 & -22.57 & 23 & 38 \\
             NGC6497 & 863 & 0.010 & 112 & 0.51 & Sab & 10.89 & -22.09 & 13 & 34 \\
             NGC6515 & 864 & 0.023 &  12 & 0.35 & E3 & 15.60 & -22.73 & 19 & 28 \\
             NGC6762 & 867 & 0.010 & 119 & 0.72 & Sab &  2.42 & -20.46 &  9 & 31 \\
             NGC6941 & 869 & 0.021 & 131 & 0.26 & Sb &  8.77 & -22.39 & 20 & 32 \\
             NGC6945 & 870 & 0.013 & 127 & 0.36 & S0 & 24.49 & -21.91 & 13 & 31 \\
             NGC6978 & 871 & 0.020 & 126 & 0.57 & Sb & 10.79 & -22.15 & 18 & 34 \\
             NGC7025 & 874 & 0.017 &  39 & 0.32 & S0a & 33.65 & -22.73 & 13 & 31 \\
             NGC7047 & 876 & 0.019 & 107 & 0.45 & Sbc &  6.18 & -21.83 & 18 & 29 \\
             NGC7194 & 881 & 0.027 &  18 & 0.30 & E3 & 27.86 & -23.05 & 17 & 22 \\
             NGC7311 & 886 & 0.015 &   9 & 0.47 & Sa & 11.72 & -22.45 & 12 & 37 \\
             NGC7321 & 887 & 0.024 &  14 & 0.32 & Sbc &  8.53 & -22.48 & 15 & 32 \\
             NGC7364 & 889 & 0.016 &  65 & 0.32 & Sab &  7.62 & -22.04 & 12 & 32 \\
            NGC7436B & 893 & 0.025 &  41 & 0.15 & E2 & 82.04 & -23.50 & 27 & 27 \\
             NGC7466 & 896 & 0.025 &  25 & 0.62 & Sbc &  5.60 & -21.86 & 13 & 31 \\
             NGC7489 & 898 & 0.021 & 160 & 0.47 & Sbc &  3.17 & -22.07 & 20 & 39 \\
             NGC7549 & 901 & 0.016 &  16 & 0.60 & Sbc &  3.97 & -21.75 & 20 & 34 \\
             NGC7550 & 900 & 0.017 & 154 & 0.09 & E4 & 27.04 & -22.89 & 24 & 25 \\
             NGC7562 & 903 & 0.012 &  83 & 0.32 & E4 & 17.66 & -22.54 & 20 & 36 \\
             NGC7563 & 902 & 0.014 & 149 & 0.47 & Sa &  9.18 & -21.54 &  9 & 31 \\
             NGC7591 & 904 & 0.017 & 150 & 0.46 & Sbc &  5.75 & -21.91 & 16 & 33 \\
             NGC7608 & 907 & 0.012 &  18 & 0.73 & Sbc &  1.24 & -20.00 & 20 & 33 \\
             NGC7611 & 908 & 0.011 & 134 & 0.55 & S0 &  7.93 & -21.32 & 11 & 21 \\
             NGC7619 & 911 & 0.013 &  50 & 0.17 & E3 &  8.79 & -22.69 & 35 & 34 \\
             NGC7623 & 912 & 0.012 &   7 & 0.30 & S0 &  9.57 & -21.47 & 10 & 31 \\
             NGC7625 & 913 & 0.005 &  10 & 0.04 & Sa &  1.33 & -20.26 & 14 & 34 \\
             NGC7631 & 914 & 0.013 &  76 & 0.62 & Sb &  3.38 & -21.10 & 17 & 33 \\
             NGC7653 & 915 & 0.014 & -11 & 0.18 & Sb &  3.16 & -21.58 & 12 & 38 \\
             NGC7671 & 916 & 0.013 & 133 & 0.37 & S0 &  9.04 & -21.76 & 11 & 26 \\
             NGC7683 & 917 & 0.012 & 138 & 0.48 & S0 & 10.45 & -21.74 & 14 & 33 \\
             NGC7684 & 919 & 0.017 &  22 & 0.66 & S0 &  9.68 & -21.69 &  9 & 38 \\
             NGC7691 & 920 & 0.013 & 171 & 0.21 & Sbc &  1.64 & -21.34 & 28 & 34 \\
             NGC7711 & 923 & 0.014 &  92 & 0.55 & E7 & 11.30 & -22.02 & 15 & 42 \\
             NGC7716 & 924 & 0.009 &  31 & 0.19 & Sb &  2.45 & -21.04 & 21 & 38 \\
             NGC7722 & 925 & 0.013 & 148 & 0.27 & Sab & 17.58 & -22.05 & 21 & 24 \\
             NGC7738 & 927 & 0.023 &  34 & 0.59 & Sb & 12.00 & -22.23 & 14 & 37 \\
        NGC7783NED01 & 932 & 0.026 & 120 & 0.54 & Sa & 28.51 & -22.59 & 15 & 31 \\
             NGC7787 & 933 & 0.022 & 104 & 0.71 & Sab &  4.18 & -21.17 & 11 & 23 \\
             NGC7800 & 937 & 0.006 &  44 & 0.61 & Ir &  0.19 & -19.56 & 32 & 37 \\
             NGC7819 & 003 & 0.017 & 105 & 0.41 & Sc &  2.45 & -21.06 & 23 & 37 \\
             NGC7824 & 006 & 0.020 & 143 & 0.37 & Sab & 17.62 & -22.26 & 11 & 38 \\
            UGC00005 & 002 & 0.024 &  44 & 0.53 & Sbc &  6.78 & -22.09 & 16 & 33 \\
            UGC00029 & 004 & 0.029 & 173 & 0.30 & E1 & 10.86 & -22.66 & 17 & 13 \\
            UGC00036 & 007 & 0.021 &  18 & 0.61 & Sab & 10.05 & -21.69 & 10 & 20 \\
            UGC00148 & 012 & 0.014 &  96 & 0.75 & Sc &  1.29 & -20.75 & 20 & 36 \\
            UGC00312 & 014 & 0.014 &   7 & 0.46 & Sd &  0.60 & -20.69 & 20 & 38 \\
       UGC00335NED02 & 017 & 0.018 & 149 & 0.49 & E4 &  6.07 & -21.39 & 18 & 24 \\
            UGC00809 & 040 & 0.014 &  23 & 0.81 & Scd &  0.49 & -19.72 & 20 & 36 \\
            UGC00841 & 041 & 0.019 &  54 & 0.77 & Sbc &  1.03 & -20.26 & 17 & 31 \\
            UGC00987 & 049 & 0.016 &  30 & 0.64 & Sa &  4.11 & -21.21 & 12 & 34 \\
            UGC01057 & 053 & 0.021 & 152 & 0.69 & Sc &  1.27 & -20.81 & 14 & 27 \\
            UGC01271 & 059 & 0.017 &  99 & 0.47 & S0a &  6.71 & -21.42 &  9 & 29 \\
            UGC02222 & 103 & 0.017 &  96 & 0.57 & S0a &  5.53 & -21.42 & 10 & 23 \\
            UGC02229 & 104 & 0.024 & 177 & 0.52 & S0a &  7.76 & -22.03 & 19 & 25 \\
            UGC02403 & 115 & 0.014 & 153 & 0.59 & Sb &  3.19 & -20.80 & 19 & 26 \\
            UGC03151 & 135 & 0.015 &  93 & 0.73 & Sa &  5.71 & -21.41 & 20 & 30 \\
            UGC03253 & 146 & 0.014 &  87 & 0.47 & Sb &  2.69 & -21.16 & 15 & 33 \\
            UGC03539 & 148 & 0.011 & 117 & 0.69 & Sc &  0.70 & -19.69 & 20 & 38 \\
            UGC03899 & 150 & 0.013 &  44 & 0.70 & Sd &  0.17 & -19.23 &  9 & 30 \\
            UGC03944 & 152 & 0.013 & 120 & 0.57 & Sbc &  0.99 & -20.42 & 17 & 33 \\
            UGC03969 & 153 & 0.027 & 134 & 0.78 & Sb &  4.78 & -21.19 & 15 & 29 \\
            UGC03995 & 155 & 0.016 &  90 & 0.56 & Sb &  8.36 & -22.12 & 25 & 39 \\
            UGC04029 & 157 & 0.015 &  63 & 0.79 & Sc &  2.14 & -20.75 & 26 & 37 \\
            UGC04132 & 165 & 0.017 &  27 & 0.69 & Sbc &  5.82 & -21.81 & 22 & 35 \\
            UGC04145 & 167 & 0.016 & 138 & 0.53 & Sa &  9.10 & -21.43 &  9 & 29 \\
            UGC04197 & 174 & 0.015 & 130 & 0.79 & Sab &  5.15 & -20.92 & 18 & 41 \\
            UGC04280 & 185 & 0.012 &   3 & 0.68 & Sb &  1.37 & -20.29 & 11 & 36 \\
            UGC04308 & 187 & 0.012 & 113 & 0.14 & Sc &  1.84 & -21.29 & 24 & 33 \\
            UGC04722 & 231 & 0.006 &  31 & 0.79 & Sdm &  0.05 & -18.18 & 32 & 38 \\
            UGC05108 & 278 & 0.027 & 138 & 0.60 & Sb &  7.74 & -22.12 &  9 & 19 \\
            UGC05113 & 281 & 0.023 &  41 & 0.74 & S0a & 12.62 & -21.76 &  8 & 22 \\
       UGC05498NED01 & 314 & 0.021 &  61 & 0.79 & Sa &  6.38 & -21.36 & 13 & 31 \\
            UGC05598 & 326 & 0.019 &  35 & 0.74 & Sb &  1.71 & -20.75 & 15 & 27 \\
            UGC05771 & 341 & 0.025 &  60 & 0.33 & E6 & 20.75 & -22.35 & 12 & 27 \\
            UGC05990 & 361 & 0.005 &  15 & 0.74 & Sc &  0.16 & -18.32 & 12 & 33 \\
            UGC06036 & 364 & 0.022 & 100 & 0.73 & Sa & 14.86 & -21.93 & 11 & 38 \\
            UGC06312 & 386 & 0.021 &  49 & 0.64 & Sab & 10.74 & -21.91 & 13 & 29 \\
            UGC07012 & 486 & 0.010 &  12 & 0.51 & Scd &  0.28 & -19.91 & 14 & 30 \\
            UGC07145 & 500 & 0.022 & 151 & 0.63 & Sbc &  2.26 & -21.14 & 16 & 32 \\
            UGC08107 & 593 & 0.028 &  53 & 0.68 & Sa & 11.64 & -22.56 & 16 & 33 \\
            UGC08231 & 606 & 0.008 &  73 & 0.66 & Sd &  0.14 & -19.28 & 19 & 33 \\
            UGC08234 & 607 & 0.027 & 133 & 0.45 & S0 & 13.65 & -22.76 &  8 & 24 \\
            UGC08733 & 657 & 0.008 &  21 & 0.44 & Sdm &  0.26 & -19.75 & 30 & 40 \\
            UGC08778 & 664 & 0.011 & 116 & 0.70 & Sb &  1.76 & -20.30 & 15 & 27 \\
            UGC08781 & 665 & 0.025 & 160 & 0.40 & Sb & 11.38 & -22.37 & 15 & 29 \\
            UGC09067 & 714 & 0.026 &  12 & 0.54 & Sbc &  3.82 & -21.85 & 14 & 28 \\
            UGC09476 & 769 & 0.011 & 132 & 0.34 & Sbc &  1.61 & -20.95 & 21 & 40 \\
            UGC09537 & 774 & 0.029 & 140 & 0.79 & Sb & 16.60 & -22.64 & 20 & 40 \\
            UGC09542 & 775 & 0.018 &  34 & 0.70 & Sc &  2.07 & -20.96 & 21 & 37 \\
            UGC09665 & 783 & 0.009 & 138 & 0.73 & Sb &  0.99 & -19.99 & 18 & 33 \\
            UGC09873 & 797 & 0.019 & 126 & 0.75 & Sb &  1.25 & -20.38 & 21 & 33 \\
            UGC09892 & 798 & 0.019 & 101 & 0.69 & Sbc &  1.98 & -20.71 & 16 & 26 \\
            UGC10097 & 814 & 0.020 & 114 & 0.18 & E5 & 28.71 & -22.73 & 14 & 27 \\
            UGC10123 & 818 & 0.013 &  53 & 0.77 & Sab &  3.32 & -20.55 & 18 & 31 \\
            UGC10205 & 822 & 0.022 & 133 & 0.38 & S0a &  9.93 & -22.32 & 19 & 35 \\
            UGC10257 & 825 & 0.013 & 162 & 0.78 & Sbc &  1.21 & -20.47 & 20 & 38 \\
            UGC10297 & 827 & 0.008 & 179 & 0.83 & Sc &  0.29 & -19.11 & 18 & 40 \\
            UGC10331 & 828 & 0.015 & 140 & 0.76 & Sc &  0.77 & -20.43 & 19 & 41 \\
            UGC10337 & 830 & 0.029 &  63 & 0.72 & Sb & 10.79 & -22.17 & 17 & 26 \\
            UGC10380 & 834 & 0.029 & 108 & 0.79 & Sb & 10.21 & -21.85 & 12 & 35 \\
            UGC10384 & 837 & 0.017 &  92 & 0.73 & Sb &  1.87 & -20.73 & 11 & 35 \\
            UGC10388 & 838 & 0.015 & 128 & 0.70 & Sa &  6.56 & -21.19 & 11 & 28 \\
            UGC10650 & 843 & 0.010 &  22 & 0.78 & Scd &  0.20 & -19.32 & 23 & 43 \\
            UGC10693 & 845 & 0.028 & 103 & 0.37 & E7 & 32.14 & -23.39 & 22 & 31 \\
            UGC10695 & 846 & 0.028 & 110 & 0.35 & E5 & 19.95 & -22.70 & 24 & 27 \\
            UGC10710 & 847 & 0.028 & 147 & 0.65 & Sb &  9.68 & -22.12 & 20 & 36 \\
            UGC10796 & 852 & 0.010 &  59 & 0.42 & Scd &  0.28 & -19.56 & 20 & 32 \\
            UGC10811 & 854 & 0.029 &  91 & 0.66 & Sb &  7.48 & -21.92 & 12 & 29 \\
            UGC10905 & 858 & 0.027 & 173 & 0.56 & S0a & 40.46 & -22.92 & 15 & 25 \\
            UGC10972 & 861 & 0.016 &  54 & 0.78 & Sbc &  2.66 & -21.22 & 24 & 34 \\
            UGC11228 & 865 & 0.019 & 178 & 0.33 & S0 & 12.39 & -22.10 & 12 & 33 \\
            UGC11649 & 872 & 0.013 &  63 & 0.22 & Sab &  3.70 & -21.38 & 19 & 32 \\
       UGC11680NED01 & 873 & 0.026 &  57 & 0.46 & Sb & 12.39 & -22.56 & 16 & 28 \\
            UGC11717 & 877 & 0.021 &  37 & 0.61 & Sab &  6.95 & -21.84 & 17 & 39 \\
            UGC12054 & 885 & 0.007 &  47 & 0.74 & Sc &  0.10 & -18.41 & 15 & 33 \\
            UGC12127 & 888 & 0.027 &   0 & 0.11 & E1 & 23.39 & -23.47 & 36 & 25 \\
            UGC12185 & 890 & 0.022 & 159 & 0.56 & Sb &  4.68 & -21.56 & 12 & 33 \\
            UGC12274 & 894 & 0.026 & 143 & 0.68 & Sa & 14.19 & -22.08 & 17 & 27 \\
            UGC12308 & 895 & 0.008 & 118 & 0.79 & Scd &  0.11 & -18.88 & 27 & 38 \\
            UGC12494 & 905 & 0.014 &  37 & 0.67 & Sd &  0.28 & -19.67 & 20 & 43 \\
            UGC12518 & 910 & 0.009 &  23 & 0.64 & Sb &  1.80 & -19.45 & 17 & 34 \\
            UGC12519 & 909 & 0.015 & 157 & 0.70 & Sc &  1.09 & -20.56 & 21 & 34 \\
            UGC12723 & 926 & 0.018 &  75 & 0.82 & Sc &  0.76 & -19.77 & 17 & 27 \\
            UGC12810 & 929 & 0.027 &  56 & 0.61 & Sbc &  5.43 & -22.01 & 20 & 35 \\
            UGC12816 & 930 & 0.018 & 140 & 0.50 & Sc &  0.66 & -20.63 & 16 & 34 \\
            UGC12857 & 934 & 0.008 &  35 & 0.72 & Sbc &  0.56 & -19.49 & 19 & 36 \\
            UGC12864 & 935 & 0.016 & 110 & 0.61 & Sc &  1.13 & -20.69 & 27 & 38 \\
          VV488NED02 & 892 & 0.016 &  70 & 0.77 & Sb &  2.32 & -20.96 & 23 & 33 \\
\enddata
\vspace{-0.8cm}
\tablecomments{Col.~1: galaxy name. Col.~2: CALIFA identification number for each galaxy. Col.~3 redshift of the galaxy from SDSS \citep{DR7}. Col.~4: position angle of the galaxy measured in the outer parts, using SDSS images. Col.~5: average ellipticity measured in the outer parts of the galaxy, using SDSS images. Col.~6: Hubble type of the galaxy from \citet{Walcher_etal_2014}. Col.~7: total stellar mass of the galaxy, measured as described in \citet{Walcher_etal_2014}. Col.~8: total absolute magnitude in $r-$band from SDSS \citep{DR7}. Col.~9: effective radii (in arcsec) of the galaxy, measured as described in \citet{Walcher_etal_2014}. Col.~10: maximum radial extent of our kinematic maps (in arcsec).}
\label{tab:sample}
\end{deluxetable}

\clearpage
\begin{appendix}

\section{Stellar kinematic maps}
\label{A:kin_maps}

This online material presents all the stellar velocity (Figs.~A1--A17) and 
velocity dispersion (Figs.~A18--A34) maps extracted from the V1200 grating used 
in this paper. The complete sample comprises 300 galaxies of Hubble 
morphological types ranging from ellipticals to late-type spirals. Velocity maps 
are in \kms\ and use a fixed range in the interval [$-150$,150] \kms. Velocity 
dispersion maps are also expressed in \kms\ and use a fixed range from 20 to 300\,\kms. 
Color schemes as in Fig.~\ref{fig:examples}. Overlaid contours come from 
SDSS $g-$band images and have been limited to the isophote reaching $\sim$2\,\Reff.
All panels cover an area of 80\arcsec$\times$100\arcsec.

\newcounter{ct}
\forloop{ct}{1}{\value{ct}<18}{

   \clearpage
   \begin{figure*}
   \centering
   \includegraphics[height=23.5cm]{califa_velocity_maps_page\arabic{ct}_v2-eps-converted-to_lowres.jpg}
   \caption{Stellar velocity maps from the CALIFA V1200 dataset.}
   \end{figure*}

}

\forloop{ct}{1}{\value{ct}<18}{

   \clearpage
   \begin{figure*}
   \centering
   \includegraphics[height=23.5cm]{califa_sigma_maps_page\arabic{ct}_v2-eps-converted-to_lowres.jpg}
   \caption{Stellar velocity dispersion maps from the CALIFA V1200 dataset.}
   \end{figure*}

}

\end{appendix}

\end{document}